\documentclass[letterpaper]{JHEP3}
\usepackage{array}
\usepackage{nicefrac}
\usepackage{amsmath,epsfig}
\usepackage{amssymb,amsfonts}

\usepackage{graphics}
\usepackage{epsfig}
\usepackage{amsfonts}

\newcommand{\Tr}{\mbox{Tr}}

\global \long \def \NN{ \mathcal{N}}

\def\g{\gamma}

\def\a{\alpha}
\def\b{\beta}
\def\e{\epsilon}

\def\h{\eta}

\def\IC{\relax\hbox{$\inbar\kern-.3em{\rm C}$}}

\def\IC{{\bf C}}

\def\bea{\begin{eqnarray}}
\def\eea{\end{eqnarray}}
\def\be{\begin{equation}}
\def\ee{\end{equation}}
\def\ba{\begin{align}}
\def\ea{\end{align}}
\def\bse{\begin{subequations}}
\def\ese{\end{subequations}}
\def\1F1{{}_1\!F_1}
\def\2F0{{}_2\!F_0}

\def\a{\alpha}

\def\h3{$\textrm{H}_3^+$}

\def\IC{{\mathbb C}}

\def\Tr{{\rm Tr}}

\def\lbldef#1#2{\expandafter\gdef\csname #1\endcsname {#2}}

\def\href#1#2{#2}

\newcommand{\beq}{\begin{equation}}
\newcommand{\eeq}{\end{equation}}
\newcommand{\ber}{\begin{eqnarray}}
\newcommand{\eer}{\end{eqnarray}}

\def\be{\begin{eqnarray}}
\def\ee{\end{eqnarray}}

\providecommand{\tabularnewline}{\\}

\def\({\left(}
\def\){\right)}
\def\[{\left[}
\def\]{\right]}
\def\<{\langle}
\def\>{\rangle}

\def\gg{\mathsf g }

\title{S-duality  and 2d Topological QFT }
\preprint{YITP-SB-09-30}
\author{Abhijit Gadde\footnote{abhijit@insti.physics.sunysb.edu},
 Elli Pomoni\footnote{pomoni@insti.physics.sunysb.edu},
 Leonardo Rastelli\footnote{leonardo.rastelli@stonybrook.edu}, and
Shlomo S. Razamat\footnote{razamat@max2.physics.sunysb.edu}
\\ 
\\
\it C.N. Yang Institute for Theoretical Physics,\\
\it Stony Brook University, \\
\it Stony Brook, NY 11794-3840, USA}

\bigskip

\abstract{

\medskip

We  study the superconformal index for the class of
 ${\mathcal N}=2$ 4d superconformal field theories recently introduced by Gaiotto \cite{Gaiotto:2009we}.
 These theories are defined by compactifying the $(2,0)$ 6d theory
 on a Riemann surface 
 with punctures.  We interpret the index of the 4d theory associated to an $n$-punctured Riemann surface
as the $n$-point correlation function of a 2d  topological QFT living on the surface. 
Invariance of the index under generalized S-duality transformations (the mapping class group of the Riemann surface)
translates into associativity of the operator algebra of the 2d TQFT.
In the $A_1$ case, for which the 4d SCFTs have a 
 Lagrangian realization, the structure constants and metric of the  2d TQFT 
can be calculated explicitly  in terms of elliptic gamma functions.
Associativity then holds thanks to a remarkable symmetry of 
 an elliptic hypergeometric beta integral, proved very recently by van de Bult \cite{Focco}.
}


\begin{document}
%\maketitle \setcounter{tocdepth}{2}
%\tableofcontents

\section{Introduction}

Electric-magnetic  duality (S-duality) in four-dimensional gauge theory has a deep connection with two-dimensional modular invariance.
The canonical example is the $SL(2, \mathbb{Z})$ symmetry of ${\cal N} = 4$ super-Yang-Mills, which can be interpreted as the modular group of a torus.
A physical picture for this correspondence is provided by the existence of the six-dimensional  $(2, 0)$ superconformal field 
theory, whose compactification on a torus of modular parameter $\tau$ yields ${\cal N} = 4$ SYM with holomorphic coupling $\tau$
(see  \cite{Witten:2009at} for a recent discussion).

Gaiotto \cite{Gaiotto:2009we} has recently discovered a beautiful generalization of this construction.
A large class of ${\cal N} = 2$ superconformal field theories in 4d is obtained by compactifying   a twisted version of the $(2, 0)$  theory on 
a Riemann surface $\Sigma$, of genus $\gg$ and  with $n$ punctures. 
The complex structure moduli space ${\cal T}_{\gg,n}/\Gamma_{\gg,n}$
of $\Sigma$ is identified with the space of exactly marginal couplings of  the 4d theory. The mapping class group $\Gamma_{\gg,n}$ acts
as the group of generalized S-duality transformations of the 4d theory.  
A striking correspondence between the Nekrasov's instanton partition function ~\cite{Nekrasov:2002qd} 
of the 4d theory and Liouville field theory on $\Sigma$ has been conjectured in \cite{Alday:2009aq} and further explored in 
\cite{Wyllard:2009hg,Drukker:2009tz,Drukker:2009id,Alday:2009fs,Mironov:2009qn,Alday:2009qq,Poghossian:2009mk,Marshakov:2009gn,
Bonelli:2009zp,Marshakov:2009kj,Nanopoulos:2009au,Gaiotto:2009ma}.
Relations to  string/M theory have been discussed in~\cite{Gaiotto:2009gz,Tachikawa:2009rb, Dijkgraaf:2009pc, Benini:2009gi}. See 
also \cite{Maruyoshi:2009uk,Nanopoulos:2009xe,Benini:2009mz}.

In this paper we study  the superconformal index~\cite{Kinney:2005ej} for this  class  of 4d SCFTs.
The index captures ``cohomological'' information about the protected
states of the theory. By construction, it counts (with signs) the protected states of the theory, up to  equivalence
 relations that set to zero  all sequences of short multiplets that may in principle recombine into long multiplets.
  
 The index is   invariant under continuous deformations of the theory, and is also expected to be invariant
under the S-duality group $\Gamma_{\gg,n}$.  {\it Assuming} $S$-duality,
this  implies that the index must be computed by 
 a topological QFT living on $\Sigma$. The usual physical arguments involving the $(2, 0)$ theory give 
 a ``proof'' of this assertion, as follows. The index has a path integral representation \cite{Kinney:2005ej}
 as the partition function of the 4d theory on $S^3 \times S^1$, twisted by various chemical potentials,
 which uplifts to a (suitably twisted) path integral of the $(2,0)$ theory on $S^3 \times S^1 \times \Sigma$. This path integral must be
independent of the metric on $\Sigma$. In the limit of small $\Sigma$  we recover the 4d definition;
in the opposite limit of large $\Sigma$ we expect a purely 2d description. Each puncture on $\Sigma$
should be regarded as an operator insertion.  By this logic, the index must be equal to the $n$-point correlation function
of some TQFT on $\Sigma$. The question is whether one can describe this TQFT more directly, and in the process {\it check}
 the S-duality of the index. 
 
 It is likely that a ``microscopic'' Lagrangian formulation of the 2d TQFT may be derived from the dimensional
 reduction of the twisted $(2,0)$ theory that we have just described, but we will not pursue this here.  Our  approach will be to start with the 4d definition
 of the index \cite{Kinney:2005ej} and write its concrete expression for Gaiotto's $A_1$ theories, 
 which have a 4d Lagrangian description. We  show  in section \ref{setup} that  the index does indeed take  the form expected
 for a correlator in a 2d TQFT. We  then
 evaluate explicitly the structure constants and metric of the TQFT operator algebra,
 and check its associativity, which is the 2d counterpart of S-duality (section \ref{secass}). The metric and structure constants have
 elegant expressions in terms of elliptic Gamma functions and the index in terms of elliptic Beta integrals,  a set of special functions
 which  are a new and active
 branch of mathematical research, see {\it e.g.} \cite{Spiridonov,Spiridonov2,Spiridonov3} and references therein. 
 For Gaiotto's $A_1$ theories associativity of the topological algebra (and thus S-duality) hinges on the invariance
 of a special case of the $E^{(5)}$ elliptic Beta integral under the Weyl group of $F_4$. A proof of this symmetry appeared on the math ArXiv 
 just as this paper was nearing completion \cite{Focco}.\footnote{We are grateful to Fokko J. van de Bult for sending us a draft of \cite{Focco}  prior to publication.}
In a related physical context, elliptic identities have been used in \cite{Dolan:2008qi} (following \cite{Romelsberger:2007ec}) 
 to prove equality of the superconformal index for Seiberg-dual pairs of ${\cal N} = 1$ gauge theories.
 
 It is also natural to ask how things work for the original paradigm of a theory exhibiting S-duality, namely ${\cal N} = 4$ SYM.
 From the viewpoint of the superconformal index the only non-trivial ${\cal N} = 4$ dual pairs are  the theories based on $SO(2n+1)/Sp(n)$ gauge groups.
We study these cases in Appendix \ref{n4app}. We write integral expressions for the index of two dual theories and check their equality
``experimentally'', for the first few orders in a series expansion in the chemical potentials. It would be nice to find an analytic proof.

\vspace{0.5cm}

\begin{figure}[htbp]
\begin{center}
$\begin{array}{c@{\hspace{0.65in}}c}
\epsfig{file=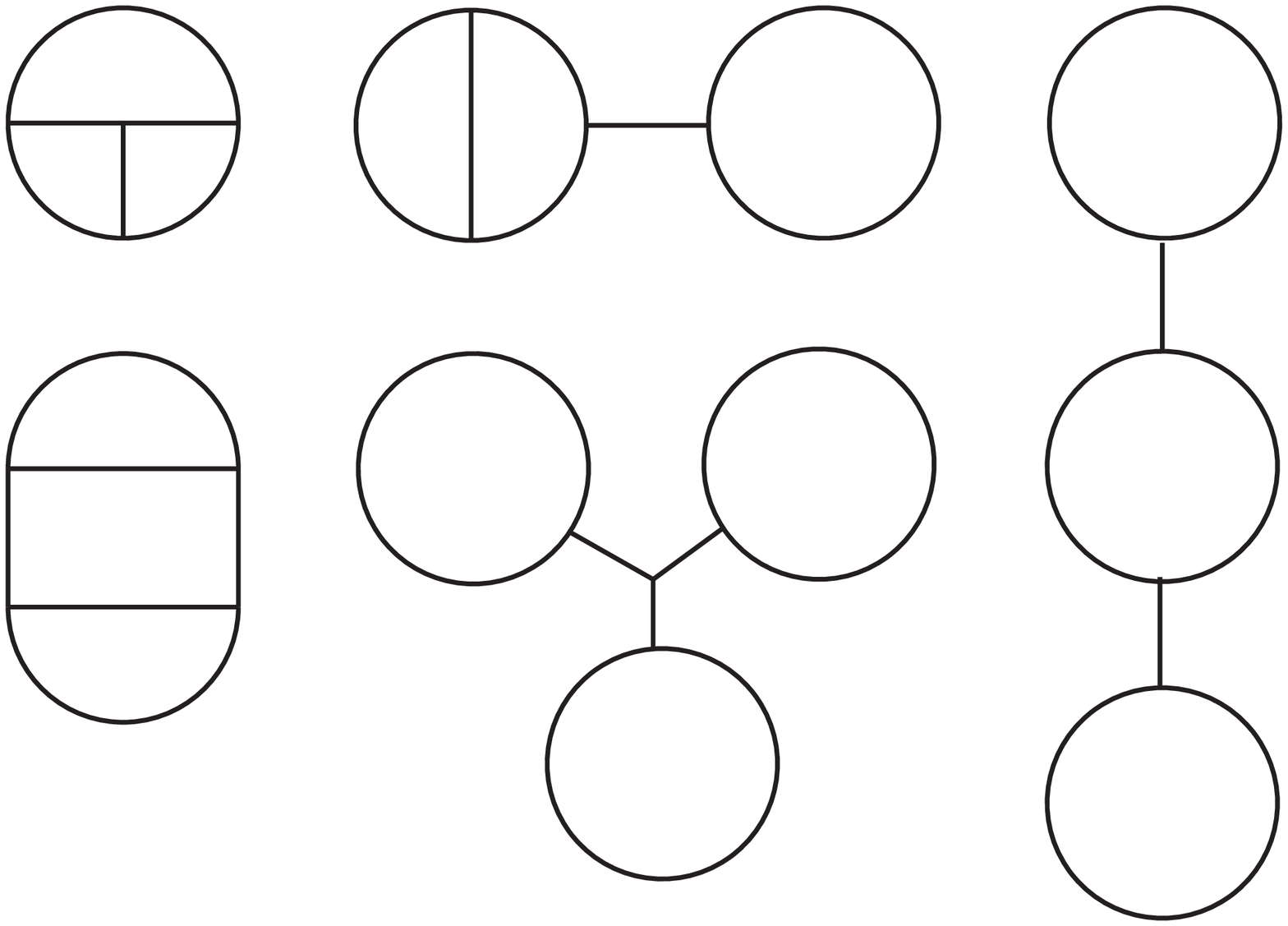,scale=0.30} & \epsfig{file=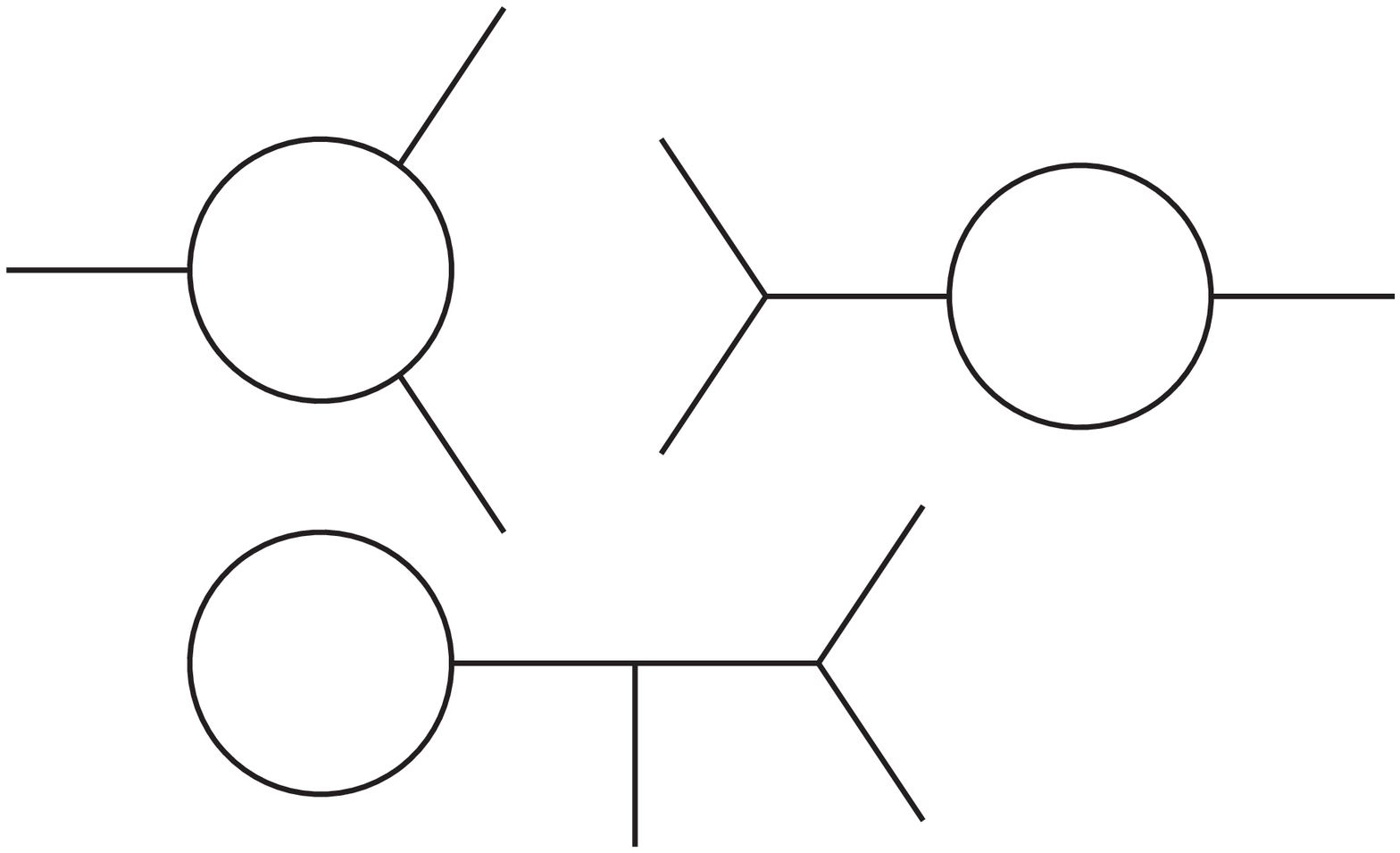,scale=0.4}\\\
(a) & (b) \\ [0.2cm]
\end{array}$
\end{center}
 \begin{center}
\caption{{\bf{(a)}} Generalized quiver diagrams representing ${\cal N} = 2$ superconformal theories with gauge group $SU(2)^6$ 
and no flavor symmetries ($N_G =6$, $N_F = 0$). There are five different theories of this kind. The internal lines of a diagram represent and $SU(2)$ gauge group
and the trivalent vertices the trifundamental chiral matter. 
{\bf{(b)}} Generalized quiver diagrams for $N_G =3$, $N_F = 3$.  Each external
leg represents an  $SU(2)$ flavor group. The upper left diagram corresponds
the ${\mathcal N}=2$ ${\mathbb Z}_3$ orbifold of ${\mathcal N}=4$ SYM with gauge group $SU(2)$. 
} \label{basic}
\end{center}
\end{figure}
\begin{figure}[htbp]
\begin{center}
$\begin{array}{c@{\hspace{1.1in}}c}
\epsfig{file=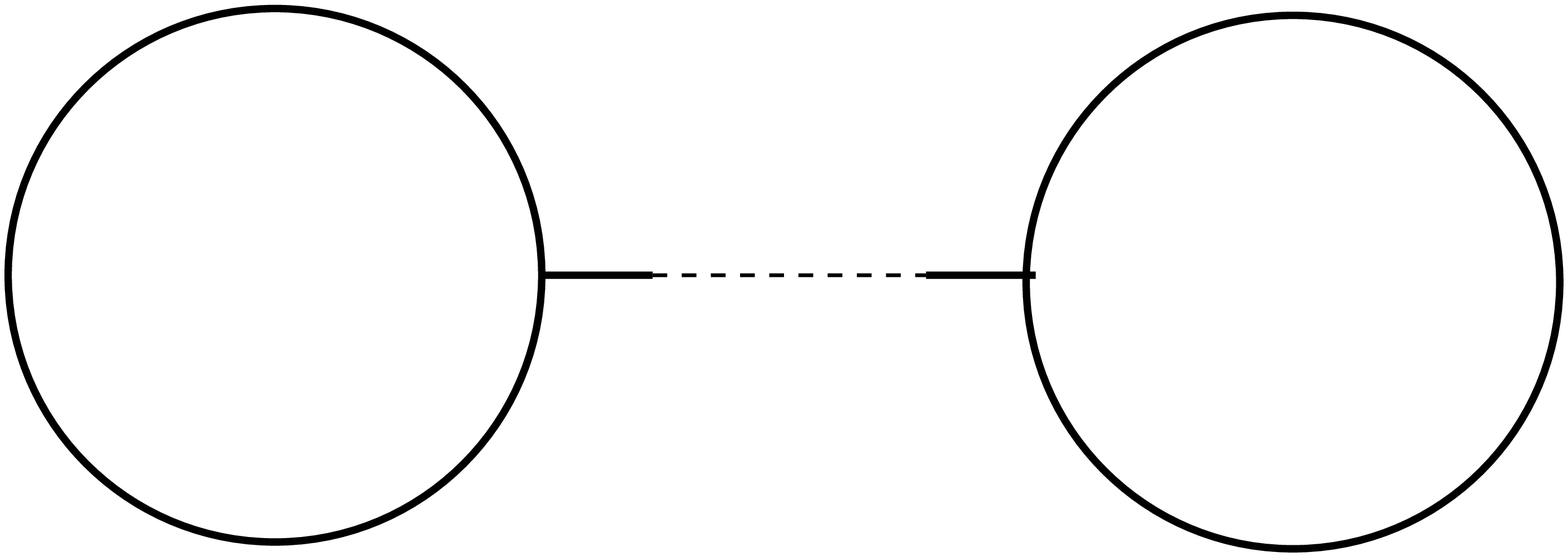,scale=0.2} & \epsfig{file=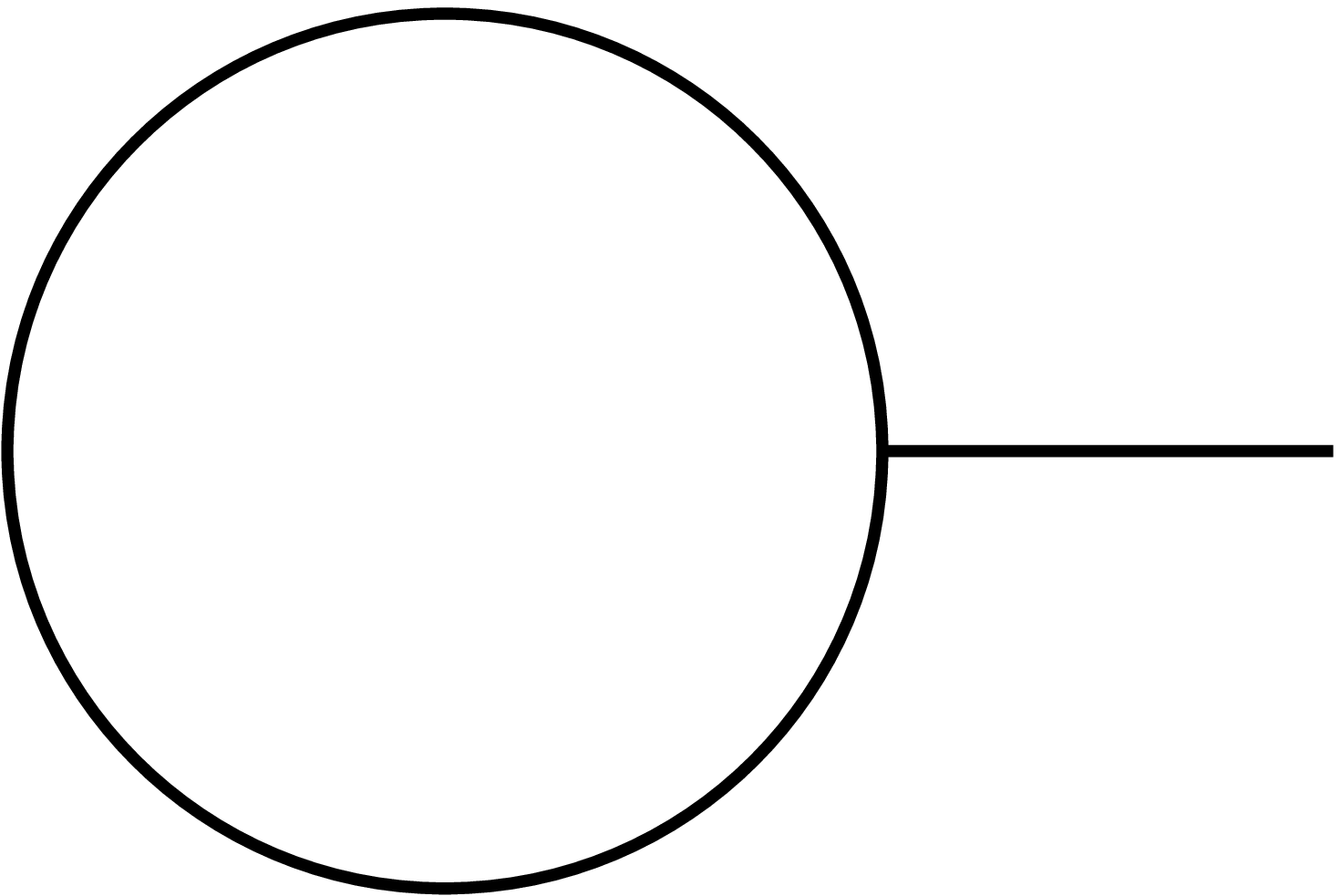,scale=0.2}\\
(a) & (b)\\ [0.2cm]
\end{array}$
\end{center}
 \begin{center}
\caption{An example of a degeneration of a graph and appearance of flavour punctures. As one of the gauge coupling is taken to  zero the corresponding
edge becomes very long. 
Cutting the edge, 
 each of the two resulting semi-infinite open legs will  be associated to chiral matter in an $SU(2)$ flavor representation.
In this picture setting the coupling 
of the middle legs in {\bf (a)} to zero gives two copies of the theory represented in {\bf (b)},
namely an $SU(2)$ gauge theory with a chiral field in the bifundamental representation 
of the gauge group  and in the fundamental of a flavour $SU(2)$. 
 } \label{degen}
\end{center}
\end{figure}

We end this introduction by recalling the basics of Gaiotto's analysis \cite{Gaiotto:2009we}.
The main achievement of \cite{Gaiotto:2009we} is a  purely four-dimensional construction of the SCFT implicitly defined by compactifying
 the $A_{N-1}$ $(2,0)$ theory on $\Sigma$.
In the $A_1$ case an explicit Lagrangian description is available, in terms of a generalized quiver with gauge group $SU(2)^{N_G}$,
see Figure \ref{basic}   for examples. The internal edges of a diagram 
correspond to the $SU(2)$ gauge groups,  the external legs to $SU(2)$ flavor groups and the
the cubic vertices  to chiral fields in the trifundamental representation (fundamental under each of the groups
joining at the vertex). The corresponding Riemann surface is immediately pictured by thickening the lines of the graph into tubes  -- with the external
tubes assumed to be infinitely long, so that they can be viewed as punctures.
The plumbing parameters $\tau_i$ of the  tubes are identified with the holomorphic gauge couplings; the degeneration limit when the surface develops a long tube corresponds to the weak coupling limit $\tau \to + i \infty$ of the corresponding gauge group (Figure \ref{degen}).  The different patterns of 
degenerations (pair-of-pants decompositions)
 of a surface $\Sigma$ of genus $\gg$ and  $N_F$ punctures
give rise to the different connected diagrams with $N_F$ external legs ($SU(2)$ flavor groups) and $N_G = N_F + 3 (\gg-1)$ internal lines ($SU(2)$ gauge groups).
Since the mapping class group permutes the diagrams,   the associated field theories must be related by generalized S-duality transformations~\cite{Gaiotto:2009we}. 

In the higher $A_{N-1}$ cases the 4d theories are generically described by more complicated quivers that  involve
 new exotic isolated SCFTs as elementary building blocks.
 While  the correspondence between the index and 2d TQFT is general, in this paper
 we will focus on the $A_1$ theories, where explicit calculations can be easily performed.

\section{2d TQFT from the Superconformal Index }
\label{setup}

The superconformal index is defined as~\cite{Kinney:2005ej}
 \be
 \mathcal{I}= {\cal I}^{WR}=  \mbox{Tr}(-1)^{F}t^{2(E+  j_2)}y^{2\,j_1}v^{-(r+R)} \,,
 \ee 
 where the trace is over the states of the theory on $S^3$ (in the usual radial quantization). 
For definiteness we are considering the ``right-handed'' Witten index ${\cal I}^{WR}$ of \cite{Kinney:2005ej}, which computes
 the cohomology of the supercharge $\bar {\cal Q}_{2 +}$, in notations \cite{Dolan:2002zh}
 where the supercharges are denoted
 as ${\cal Q}^I_\alpha$, $\bar {\cal Q}_{I  \dot \alpha}$, ${\cal S}_{I \alpha}$, $\bar {\cal S}^{I}_  {\dot \alpha}$,
 with $I =1,2$ $SU(2)_R$ indices and $\alpha=\pm$, $\dot \alpha = \pm$ Lorentz indices. (For the class of superconformal theories that we consider, the
  left-handed and right-handed Witten indices are equal.) The chemical potentials $t$, $y$, and $v$
 keep track of various combinations of quantum numbers associated to the supercorformal algebra $SU(2,2 |2)$: 
 $E$ is the conformal dimension, $(j_1, j_2)$ the $SU(2)_1 \times SU(2)_2$  Lorentz spins, and $(R \, ,r)$ the quantum numbers
 under the  $SU(2)_R \times U(1)_r$ R-symmetry.\footnote{Our normalization for the R-symmetry charges is as in \cite{Dolan:2002zh}
and  differs from \cite{Kinney:2005ej}: $R_{here} = R_{there}/2$, $r_{here} = r_{there}/2$.}
 
 For the $A_1$ generalized quivers
the index can be explicitly computed as a matrix integral,
\be
\label{index} 
{\cal I} =\int\prod_{\ell \in {\cal G} }\left[dU_\ell\right]\, \exp\left(\sum_{n=1}^\infty\frac{1}{n}\,\left[\sum_{i \in {\cal G} } f_n  \cdot \chi_{adj}(U_i^n)+\sum_{(i,j,k) \in {\cal V}} g_n  \cdot \chi_{3f}(U_i^n,
U_j^n, U_k^n \, 
)\right]\right) \, .
\ee 
Here
 $f_n = f(t^n, y^n, v^n)$ and $g_n = g(t^n, y^n, v^n)$, with $f(t, y, v)$ and $g(t, y, v)$  the ``single-letter
partition functions'' for respectively the adjoint and trifundamental degrees of freedom, multiplying
the corresponding $SU(2)$ characters. The explicit expressions for $f$ and $g$ will be given in the next section.
The $\{ U_i \}$ are $SU(2)$ matrices. Their index $i$ run over the $N_G + N_F$ edges of the diagram, both  internal 
(``Gauge'') and external (``Flavor''). 
The set $\cal G$ is the set of  $N_G$ internal edges 
 while the set  $\cal V$
is the set of trivalent vertices, each vertex being labelled by the triple $(i,j,k)$ of incident edges.
The integral over $\{ U_\ell \, , \ell \in {\cal P} \}$, with  $\left[dU\right]$ being the Haar measure, enforces the gauge-singlet condition.
 All in all, the index ${\cal I}$
depends on the chemical potentials $t$, $y$, $v$  (through $f$ and $g$) and on (the eigenvalues of) the $N_F$ unintegrated flavor matrices.

The characters depend on a single angular
variable $\alpha_i$ for each  $SU(2)$ group $U_i$.  Writing 
\be \label{alphalabel}
U_i=V_i^\dagger\,\left(\begin{array}{cc}e^{i\alpha_i}&0\\0&e^{-i\alpha_i}\end{array}\right)\,V_i \,,\qquad
\ee
we have
\begin{eqnarray}
&& \chi_{adj}(U_i) =  \Tr U_i\, \Tr U_i-1=e^{2i\a_i}+e^{-2i\a_i}+1 \equiv \chi_{adj}(\a_i) \, ,\\
&& \chi_{3f}(U_i,U_j,U_k)  =  \Tr U_i\, \Tr U_j\, \Tr U_k =(e^{i\a_i}+e^{-i\a_i})(e^{i\a_j}+e^{-i\a_j})
 (e^{i\a_k}+e^{-i\a_k})  \\  && \qquad  \qquad \qquad \qquad \qquad \qquad \qquad \; \; \, \equiv \chi_{3f}(\a_i,\a_j,\a_k)\, ,\nonumber
\end{eqnarray}
where 
we have used the fact that $2~\sim \bar 2$.  Integrating over $V_i$, the Haar measure simplifies to
\be \label{measure}
\int\left[dU_i\right]=\frac{1}{\pi}\int_0^{2\pi}d\alpha_i \;\sin^2 \alpha_i \equiv \int d\alpha_i \, \Delta(\alpha_i) \, .
\ee
We now define 
\be\label{consmet}
C_{\a_i\a_j\a_k}& \equiv &\exp\left(\sum_{n=1}^\infty\frac{1}{n}\,g_n\cdot\;\chi_{3f}(n\a_i,n\a_j,n\a_k)\right),\\
\label{etadef}
\eta^{\a_i\a_j}& \equiv &\exp\left(\sum_{n=1}^\infty\frac{1}{n}\,f_n\cdot\; \chi_{adj}(n\a_i)\right)\,\hat\delta(\a_i,\a_j)\equiv \eta^{\a_i}\,\hat\delta(\a_i,\a_j),\nonumber 
\ee 
where $\hat \delta(\a,\b) \equiv \Delta^{-1}(\alpha) \delta(\alpha-\beta)$ (with the understanding
that $\alpha$ and $\beta$ are defined modulo $2 \pi$)   is the  delta-function with respect to the measure (\ref{measure}).
Further define the  ``contraction'' of   an upper and a lower $\alpha$ labels as
\be \label{contraction}
A^{\dots \a\dots}\,B_{\dots \a\dots}\equiv \int_0^{2 \pi}d\a\, \Delta(\a)\,A^{\dots \a\dots}\,B_{\dots \a\dots} \, .
\ee
The superconformal index \eqref{index} can  then be suggestively written as
\be\label{topindex}
{\mathcal I}=\prod_{\{i,j,k\}\in {\mathcal V}} C_{\a_i\a_j\a_k}\,\prod_{\{m,n\}\in {\mathcal G}}\eta^{\a_m\a_n} \, .
\ee 
The  internal labels $\{ \alpha_i \; ,i \in {\cal G} \}$ associate to the gauge groups  are contracted,  while the $N_F$ external
labels associated to the flavor groups are left open.  The expression (\ref{topindex})
is naturally interpreted as an $N_F$-point ``correlation function'' $\langle \alpha_1 \dots \alpha_{N_F} \rangle_\gg$,
evaluated by regarding the generalized quiver as a ``Feynman diagram''.  The Feynman rules
assign to each trivalent vertex  the cubic coupling $C_{\alpha \beta \gamma}$, and to each internal propagator the inverse metric $\eta^{\alpha \beta}$.
S-duality implies that the superconformal indices calculated from two diagrams with the same $(N_F, N_G)$  must be equal. These properties
can be summarized in
the  statement  that  the superconformal index is evaluated
by a 2d Topological QFT (TQFT).

\begin{figure}[htbp]
\begin{center}
$\begin{array}{c@{\hspace{1.0in}}c}
\epsfig{file=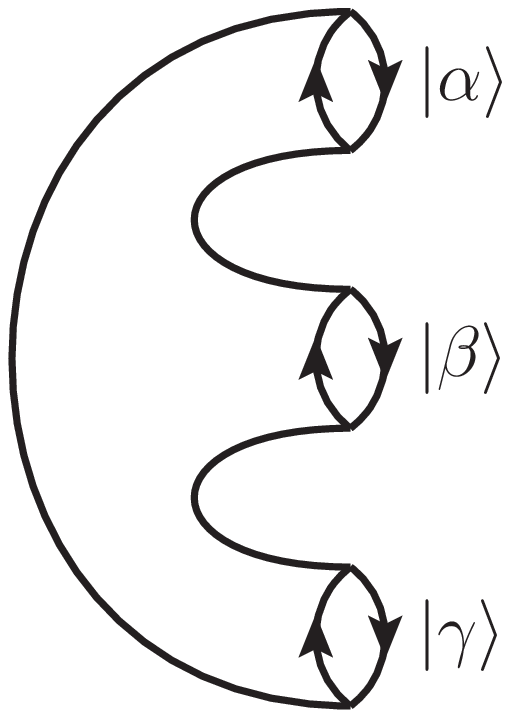,scale=0.4} & \epsfig{file=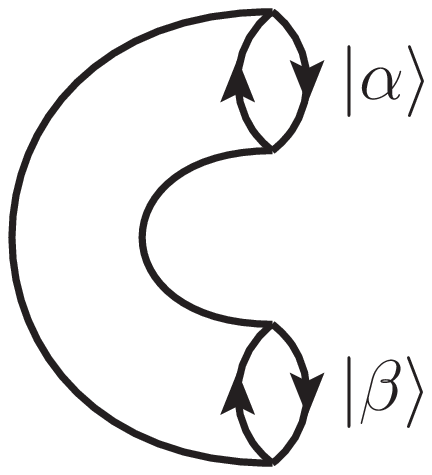,scale=0.4}\\
(a) & (b)\\ [0.2cm]
\end{array}$ 
\end{center}
\caption{{\bf (a)} Topological interpretation of the structure constants ${C_{\a\b\g}} \equiv \langle C | \, |\alpha \rangle |\beta \rangle |\gamma \rangle$. 
The path integral over the sphere with three boundaries defines  $\langle C | \in {\cal H}^* \otimes {\cal H}^* \otimes {\cal H}^*$.
{\bf (b)} Analogous interpretation of the metric $\eta_{\a\b} \equiv \langle \eta | |\alpha \rangle |\beta \rangle$, with $\langle \eta | \in {\cal H}^* \otimes {\cal H}^*$,
 in terms of the sphere with two boundaries.
}
\label{topquants} 
\end{figure}
\begin{figure}[htbp]
\begin{center}
$\begin{array}{c@{\hspace{1.0in}}c}
\epsfig{file=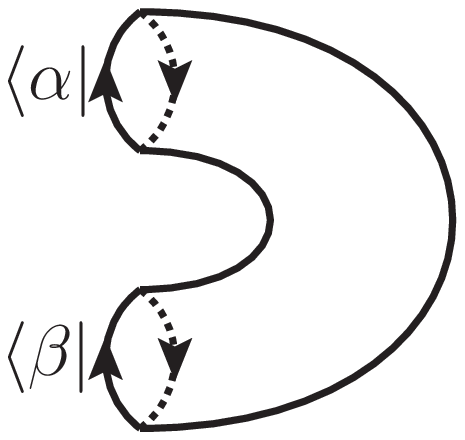,scale=0.4} & \epsfig{file=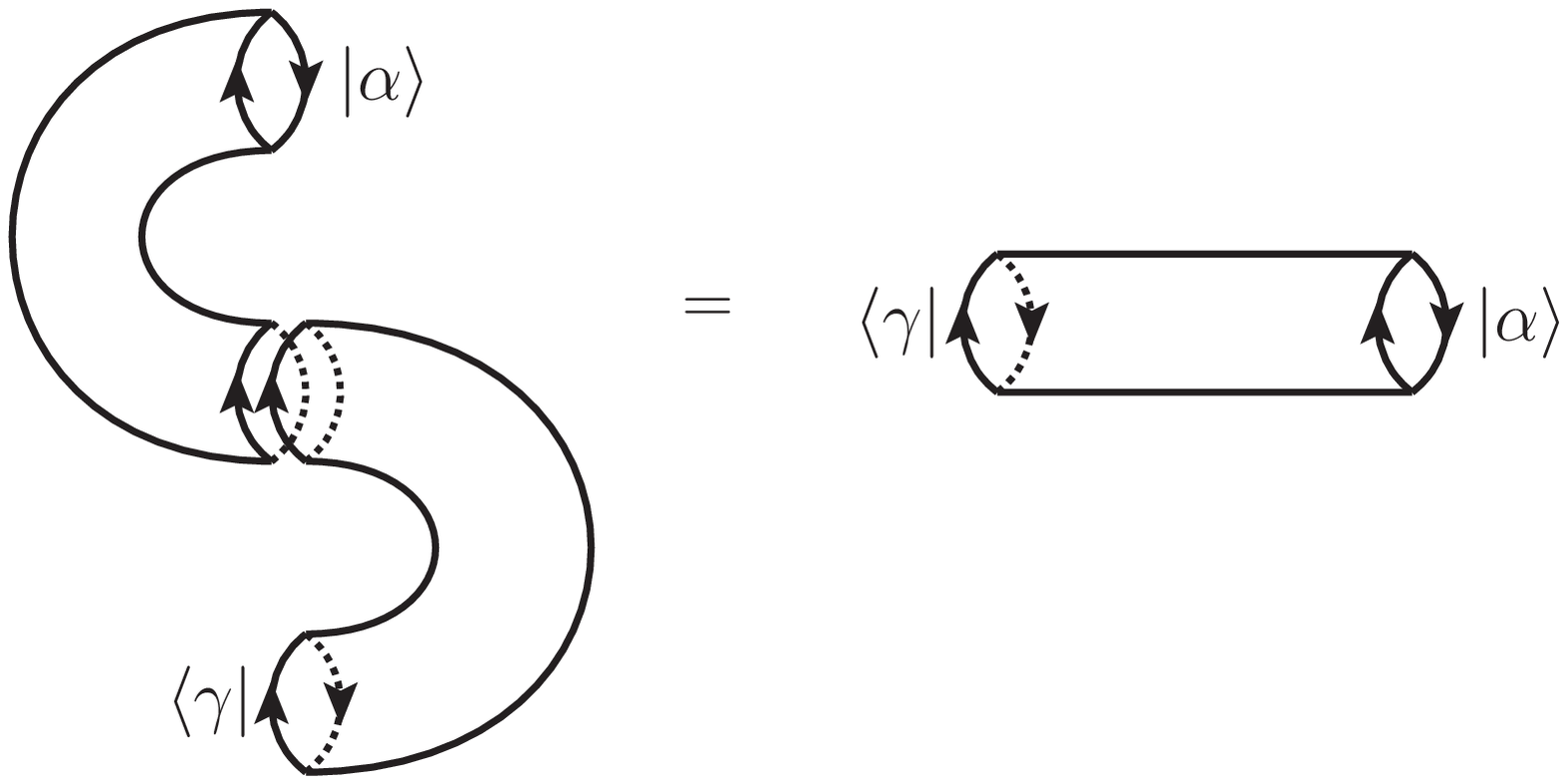,scale=0.4}\\
(a) & (b)\\ [0.2cm]
\end{array}$
\end{center}
\caption{Topological interpretation of {\bf (a)} the  inverse metric ${\eta^{\a\b}}$, {\bf (b)}
the relation $\eta_{\a\b} \eta^{\b\g}=\delta^{\g}_{\a}$.
By convention, we draw the boundaries  associated with  upper indices  facing left
and the boundaries associated with the lower indices  facing right.
}
\label{inversemetrics}
\end{figure}

At the informal level sufficient for our discussion, a 2d TQFT  \cite{atiyah,Dijkgraaf:1997ip} can be characterized in terms of the following data: a space of states ${\cal H}$;  
a non-degenerate, symmetric metric $\eta$:  ${\cal H} \otimes {\cal H} \to \mathbb{C}$;
and a completely symmetric triple product $C$: ${\cal H} \otimes {\cal H}  \otimes {\cal H} \to \mathbb{C}$. 
The states  in ${\cal H}$ are understood physically as  wavefunctionals of  field configurations on the ``spatial'' manifold $S^1$.
The metric and triple product are evaluated by  the path integral over field configurations on the sphere with respectively two and three boundaries (Figure \ref{topquants}).
The 2d surfaces where the TQFT is defined are assumed to be oriented, so the  $S^1$ boundaries inherit a canonical
orientation. To a boundary of inverse orientation (with respect to the canonical one) is associated
the {\it dual} space ${\cal H}^*$. Choosing a basis for ${\cal H}$, we can specify the metric and triple product in terms of $\eta_{\alpha \beta} \equiv \eta(|\alpha\rangle, |\beta \rangle)$ and
$C_{\alpha \beta \gamma} \equiv C(|\alpha \rangle, |\beta \rangle, |\gamma \rangle)$, or
\be
\eta = \sum_{\alpha, \beta} \eta_{\alpha \beta} \langle \alpha| \langle \beta| \, , \quad C = \sum_{\alpha, \beta, \gamma} C_{\alpha \beta \gamma} \langle \alpha| \langle \beta| \langle \gamma| \,.
\ee
The inverse metric $\eta^{\alpha \beta}$ is associated to the sphere with two boundaries of inverse orientation, and as its name
suggests it obeys $\eta^{\alpha \beta} \eta_{\beta \gamma} = \delta^\alpha_\gamma$, see Figure \ref{inversemetrics}.
Index contraction corresponds geometrically to gluing of $S^1$ boundary of compatible orientation.  

\begin{figure}[htbp]
 \begin{centering}
 \includegraphics[scale=0.4]{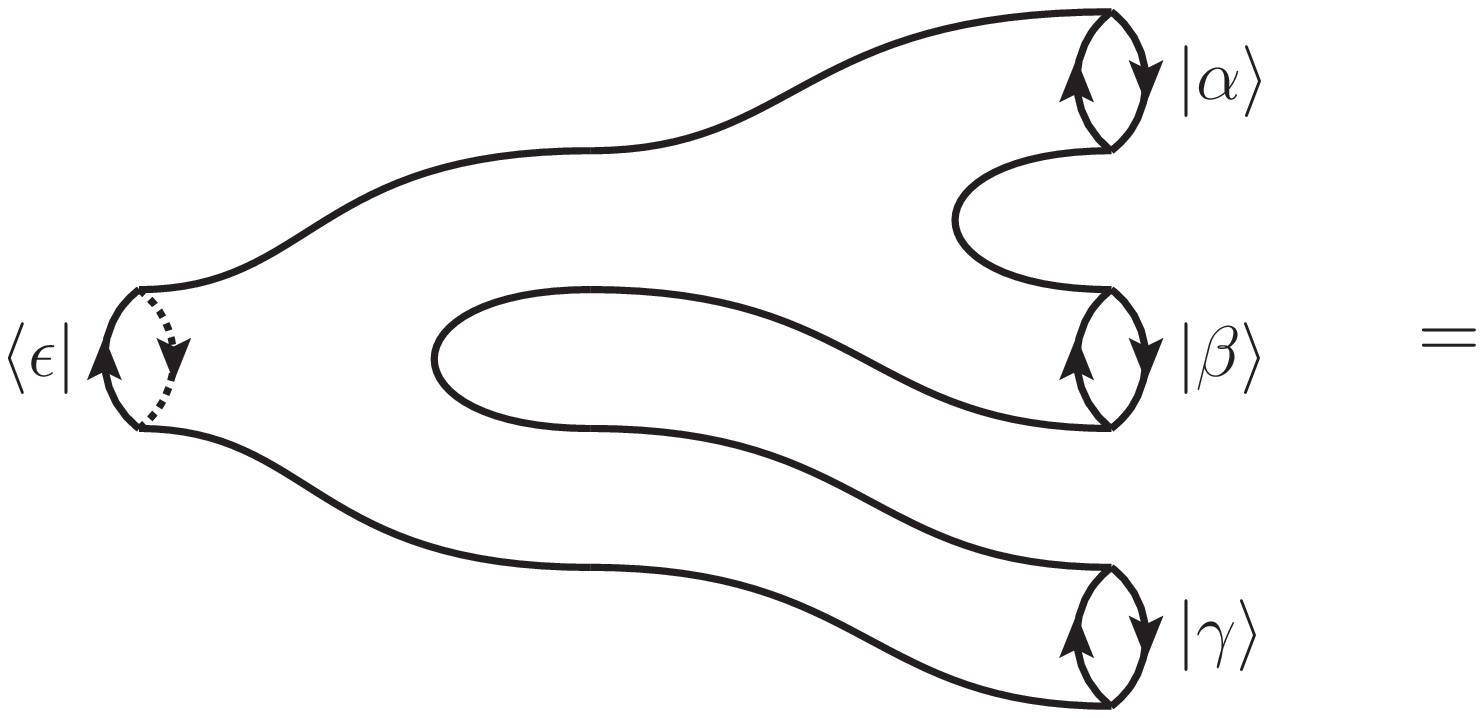} \includegraphics[scale=0.4]{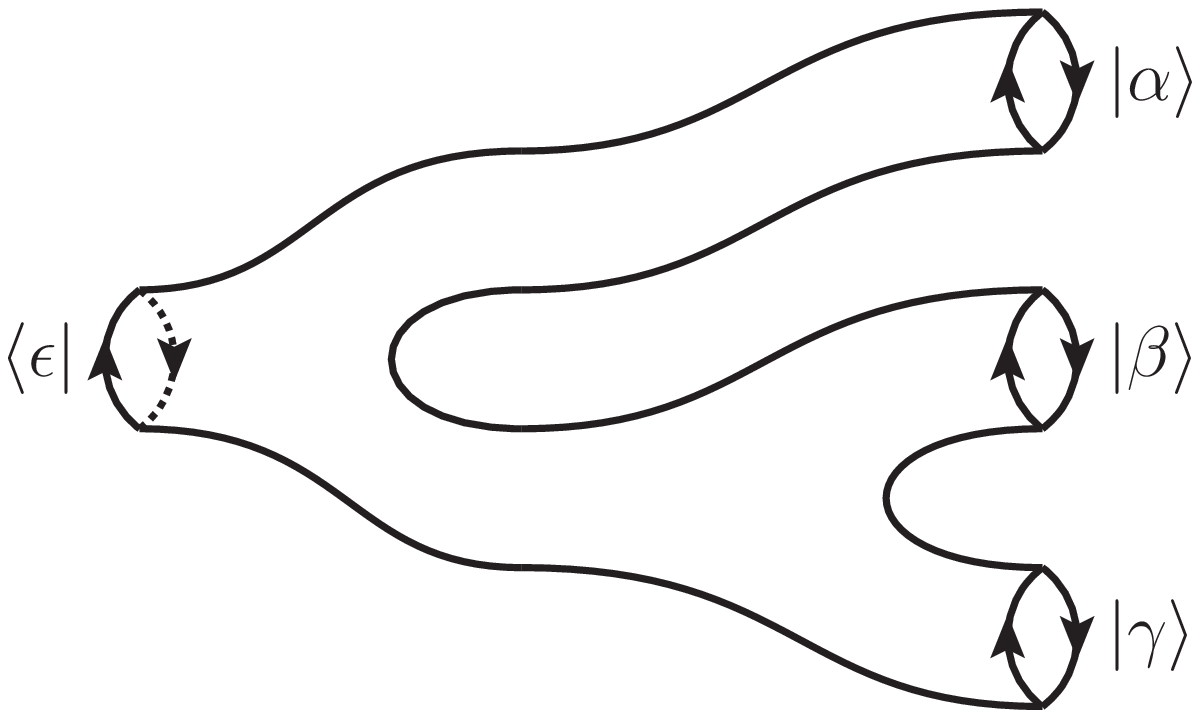}
 \par \end{centering}
 \caption{Pictorial rendering of the associativity of the algebra.}\label{assoc}
 \end{figure}

The metric and triple product obey natural compatibility
axioms which can be simply summarized by the statement that
the metric and its inverse are  used to lower and raise indices in the usual fashion. 
Finally the crucial requirement: the structure constants ${C_{\a\b}}^\gamma
\equiv C_{\alpha \beta \epsilon} \eta^{\epsilon \gamma}$ define an
associative algebra
\be\label{ax3}
{C_{\a\b}}^\delta\, {C_{\delta\g}}^\e={C_{\b\g}}^\delta\, {C_{\delta\a}}^\e \, ,
\ee
as illustrated in Figure \ref{assoc}.  From these data, arbitrary $n$-point correlators  on a genus $\gg$ surface
can be evaluated by factorization (= pair-of-pants decomposition of the surface). The result is guaranteed to be independent
of  the specific decomposition.

In our case the space  ${\cal H}$ is spanned by the states $\{ |\alpha \rangle \, ,\alpha \in [0, 2 \pi) \}$,
where $\alpha$  parametrizes the $SU(2)$  eigenvalues, equ.(\ref{alphalabel}).
Alternatively we may ``Fourier transform'' to the basis of irreducible $SU(2)$ representations, $\{ |R_K \rangle\, , K \in \mathbb{Z}_+ \}$, 
see Appendix B. We have concrete expressions (\ref{consmet}, \ref{etadef}) for the cubic couplings $C_{\alpha \beta \gamma}$ and for the {\it inverse} metric $\eta^{\alpha \beta}$,
which are manifestly symmetric under permutations of the indices. Formal inversion of (\ref{etadef}) gives the
metric $\eta_{\alpha \beta} \equiv (\eta^\alpha)^{-1} {\hat \delta}(\alpha , \beta)$. 
Finally with the help of (\ref{contraction}) 
we can raise, lower and contract indices at will.
On physical grounds we expect these formal manipulations to make sense, since the superconformal
index is well-defined as a series expansion in the chemical potential $t$, which should 
 have a finite radius of convergence \cite{Kinney:2005ej}. The explicit analysis of sections 3 and 4 will confirm these expectations.
 We will find expressions for the index as analytic
 functions of the chemical potentials.  Our definitions satisfy   the axioms  of a 2d TQFT by construction, and independently
 of the specific form of  the functions $f(t,y,v)$ and $g(t,y,v)$, except for the  associativity axiom,  which is completely non-trivial.
 Associativity of the 2d topological algebra is equivalent to 4d S-duality, and it can only hold
 for very special choices of field content, encoded in the single-letter partition functions $f$ and $g$.
   
\section{Associativity of the Algebra}\label{secass}

In this section we determine explicitly the structure constants and the metric of the TQFT and
 write them  in terms of elliptic Beta integrals. With the help of a recent mathematical result \cite{Focco} 
 we prove analytically the associativity of the topological algebra.

\subsection{Explicit Evaluation of the Index}
\label{explicit}
 \begin{table}
  \begin{centering}
  \begin{tabular}{|c|r|r|r|r|r|c|}
  \hline 
Letters & $  E$ & $j_1$ & $  j_2$ & $R$ & $r$ & $  \mathcal{I}$  \tabularnewline
  \hline
   \hline 
$  \phi$ & $1$ & $0$ & $0$ & $0$ & $-1$ & $t^{2}v$  \tabularnewline
  \hline 
$  \lambda_{\pm}^{1}$ & $  \frac{3}{2}$ & $  \pm  \frac{1}{2}$ & $0$ & $  \frac{1}{2}$ & $-  \frac{1}{2}$ & $-t^{3}\,y,\;-t^3\,y^{-1}$  \tabularnewline
  \hline 	
$  \bar{\lambda}_{2+}$  & $  \frac{3}{2}$ & $0$ & $  \frac{1}{2}$ & $  \frac{1}{2}$ & $  \frac{1}{2}$ & $-t^{4}/v$  \tabularnewline
  \hline 
$  \bar{F}_{++}$ & $2$ & $0$ & $1$ & $0$ & $0$ & $t^{6}$  \tabularnewline
  \hline
  $  \partial_{-+}  \lambda_{+}^{1}+  \partial_{++}  \lambda_{-}^{1}=0$ & $  \frac{5}{2}$ & $0$ & $  \frac{1}{2}$ & $  \frac{1}{2}$ & $  -\frac{1}{2}$ & $t^{6}$ \tabularnewline
  \hline 
\hline
$q$ & $1$ & $0$ & $0$ & $  \frac{1}{2}$ & $0$ & $t^{2}/  \sqrt{v}$  \tabularnewline
  \hline 
$  \bar{\psi}_{+}$ & $  \frac{3}{2}$ & $0$ & $  \frac{1}{2}$ & $0$ & $-  \frac{1}{2}$ & $-t^{4}  \sqrt{v}$  \tabularnewline
  \hline 
    \hline 
$  \partial_{\pm+}$ & $1$ & $  \pm  \frac{1}{2}$ & $  \frac{1}{2}$ & $0$ & $0$ & $t^{3\,}y,\;t^3\,y^{-1}$  \tabularnewline
  \hline
  \end{tabular}
  \par  \end{centering}
  \caption{Contributions to the index from  ``single letters''. 
  We denote by $(\phi, \bar \phi,  \lambda^I_\alpha, \lambda_{I \,\dot \alpha},  F_{\alpha \beta}, \bar F_{\dot \alpha \dot \beta})$
 the components of the adjoint ${\cal N} = 2$ vector multiplet,  by $(q, \bar q, \psi_\alpha, \bar \psi_{\dot \alpha})$ the components  of the trifundamental ${\cal N} = 1$
chiral multiplet,  and by $\partial_{\alpha \dot \alpha}$ the spacetime derivatives. Here $I = 1,2$ are $SU(2)_R$ indices and
$\alpha = \pm$, $\dot \alpha = \pm$ Lorentz indices.
}
\label{letters}
  \end{table}
  \begin{figure}[htbp]
\begin{center}
$\begin{array}{c}
\epsfig{file=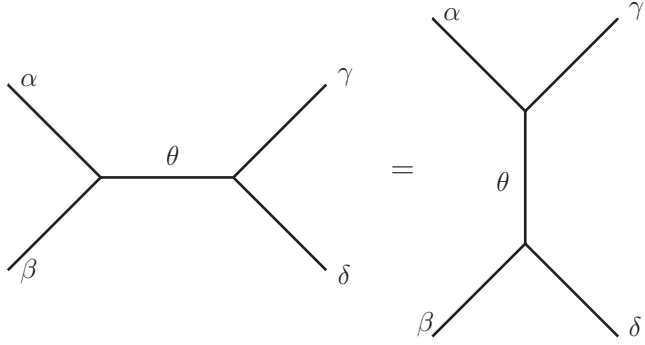,scale=0.5} 
\\ [0.2cm]
\end{array}$
\end{center}
 \begin{center}
\caption{The basic S-duality channel-crossing. The two diagrams are two equivalent (S-dual) ways to represent
the $ \NN=2$  gauge theory with a single gauge group $SU(2)$ and four $SU(2)$ flavour groups, which is the basic
 building block of the $A_1$ generalized quiver theories.  The indices on the edges label the eigenvalues of the corresponding $SU(2)$ groups.
} \label{crossing}
\end{center}
\end{figure}
The ``single letters'' contributing to the index, which must obey $\bar \Delta \equiv %2 \{ \bar Q_{2 +}, \bar S  \} =
 E - 2 j_2 - 2 R +r = 0$  \cite{Kinney:2005ej},  are enumerated in Table \ref{letters}.
The first block of the Table shows the contributing letters from the adjoint ${\cal N} = 2$ vector multiplet (associated to each internal edge of a graph),
including the equations of motion constraint. The second block shows the contributions from the ${\cal N}= 1$ chiral multiples in the trifundamental
representation, associated to each cubic vertex. Finally the last line of the Table shows the spacetime derivatives
contributing to the index. Since each field can be hit by an arbitrary number of derivatives, the derivatives give
a multiplicative contribution to the single-letter partition functions of the form
\be
\sum_{m=0}^\infty \sum_{n=0}^\infty (t^3 y)^m  \, (t^3 y^{-1})^n = \frac{1}{(1-t^3 y)(1-t^3 y^{-1})}\,.
\ee
All in all, the single letter partition function are  given by
\be\label{letterpart}
\text{adjoint}\qquad&:&\qquad f(t,y,v)=\frac{t^2v-\frac{t^4}{v}-t^3(y+y^{-1})+2t^6}{(1-t^3\,y)(1-t^3y^{-1})}\,,\\
{\rm trifundamental} \qquad&:&\qquad  g(t,y,v)=\frac{\frac{t^2}{\sqrt{v}}-t^4\sqrt{v}}{(1-t^3\,y)(1-t^3y^{-1})}\,.
\ee 

We are now ready to check explicitly  the basic S-duality move --  S-duality with respect to one of the $SU(2)$ gauge groups, 
represented graphically as channel-crossing with respect to one
of the edges of the graph (Figure \ref{crossing}). 
The full S-duality group of a graph is generated by repeated
applications of the basic move to different edges. 
The contribution to the index from the left graph in Figure \ref{crossing} is
\be
{\mathcal I}=
\int d\theta  \, \Delta(\theta)
 \exp\left(
\sum_{n=1}^\infty\frac{1}{n}\,\left[f_n  \cdot \chi_{adj}(n\theta)+g_n  \cdot \chi_{3f}(n\a, n\beta, n\theta)
+g_n  \cdot \chi_{3f}(n\theta,n\g,n\delta)\right]
\right)  \,.\nonumber\\
\ee
Substituting the expressions for the characters,
\be
\label{indexS}
{\cal I} =  \frac{e^{\sum_{n=1}^\infty\frac{f_n}{n}}}{\pi}\int_0^{2\pi} d\theta \sin^2\theta\,
 e^{ {\sum_{n=1}^\infty\frac{2f_n}{n}\cos 2n\theta} }
e^{ \sum_{n=1}^\infty\frac{8g_n}{n}\left[\cos n\a\cos n\b+\cos n\g\cos n\delta\right] \cos n\theta } \, ,
\ee where $f_n \equiv f(t^n,y^n,v^n)$ and  $g_n \equiv f(t^n,y^n,v^n)$.  
S-duality of the index is the statement this integral 
is invariant under permutations of the external labels $\a,\b,\g,\delta$. Since  symmetries under $\a \leftrightarrow \b$ and (independently)
under $\gamma \leftrightarrow \delta$ are manifest, the non-trivial requirement is symmetry under $\beta \leftrightarrow \gamma$,
which gives the index associated to the crossed graph on the right of   Figure \ref{crossing}.

 The integrand of   \eqref{indexS}  is {\it not}  invariant under  $\b\leftrightarrow\g$, but the integral is, as once  can check 
order by order in
  a series expansion in the  chemical potential $t$.
Here is how things work to the first non-trivial order.
We expand the integrand in $t$ around $t=0$, and set $y=v=1$ for simplicity.
The single-letter partition functions behave as
\be
 f(t,y=1, v=1)\sim t^2-2\,t^3 \,, \qquad g(t,y=1, v=1)\sim t^2-t^4 \, .
\ee 
 The first non-trivial check is for the coefficient of ${\cal I}$ of  order $O(t^4)$, 
\be \label{t4}
&& {\cal I} \sim t^4\;\int_0^{2\pi} d\theta \sin^2\theta\biggl(\cos 4\theta +2\cos^2 2\theta
+4 A_2\,\cos 2\theta +32A_1^2\, \cos^2\theta-\\
&&\qquad\qquad\qquad\qquad\qquad-2\cos 2\theta +16A_1\,\cos \theta \cos 2\theta -8A_1\,\cos \theta\biggr) \, ,\nonumber
\ee
where
$
A_n\equiv  \cos n\a\cos n\b+\cos n\g\cos n\delta 
$.  Performing the elementary integrals,
\be
{\cal I} \sim t^4 \left[ 6\pi  +2\pi  \;\left(\cos2\a+\cos2\b+\cos2\g+\cos2\delta+8\cos\a\cos\b\cos\g\cos\delta\right) \right]\, ,
\ee 
which is indeed symmetric under $\alpha  \leftrightarrow \beta \leftrightarrow \gamma \leftrightarrow \delta$.
We stress that crossing symmetry depends crucially on the specific
form of the single-letter partition functions \eqref{letterpart} and thus on the specific field content.
We have performed  systematic checks by calculating the series expansion to several higher orders using  {\it Mathematica}.
Fortunately it is possible to give an analytic proof of crossing symmetry of the index,
as we now describe.

\subsection{Elliptic Beta Integrals and S-duality}\label{ellapp}

The fundamental integral \eqref{indexS} can be recast in an elegant way 
in terms of special functions known as elliptic Beta integrals. 
We start by recalling the definition of the elliptic Gamma function,
 a two parameter generalization of the Gamma function,
\be\label{gamma1}
\Gamma(z;p,q) \equiv \prod_{j,k\geq 0}\frac{1-z^{-1}\,p^{j+1}q^{k+1}}{1-z\,p^jq^k} \,.
\ee 
For reviews of the elliptic Gamma function and of elliptic
hypergeometric mathematics the reader can consult~\cite{Spiridonov,Spiridonov2,Spiridonov3}.
 Throughout this paper we will use the standard condensed notations
\be\label{condensed}
&&\Gamma(z_1,\dots, z_k;p,q)\equiv \prod_{j=1}^k \Gamma(z_j;p,q),\\ 
&&\Gamma(z^{\pm1};p,q)=\Gamma(z;p,q)\Gamma(1/z;p,q) \,.\nonumber
\ee
Two identities satisfied by the elliptic Gamma function  that will be useful to us are
\be
&&\Gamma(z^2;p,q)=\Gamma(\pm z,\pm\sqrt{q}\,z,\pm\sqrt{p}\,z,\pm\sqrt{pq}\, z;p,q)\, ,\label{double}\\
&&\Gamma\left(pq/z;p,q\right)\Gamma\left(z;p,q\right)=1\,.\label{inverse} 
\ee 
(As an illustration of  the shorthand \eqref{condensed}, the {\it rhs} of (\ref{double}) is a product of eight Gamma functions.)
Using the definition (\ref{gamma1}), 
it is  straightforward to show~\cite{Dolan:2008qi} 
\be\label{dolanid} 
&&\exp\left({\sum_{n=1}^\infty \frac{1}{n}}\,\frac{t^{2n}z^n-t^{4n}z^{-n}}{(1-t^{3n} y^n)(1-t^{3n}y^{-n})}\right)=\Gamma(t^2\,z; p,q),\\
&&\exp\left({\sum_{n=1}^\infty \frac{1}{n}}\,\frac{2t^{6n}-t^{3n}(y^n+y^{-n})}{(1-t^{3n} y^n)(1-t^{3n}y^{-n})}(z^n+z^{-n})\right)=
-\frac{z}{(1-z)^2}\,\frac{1}{\Gamma(z^{\pm1};p,q)},\nonumber
\ee
where
\be
p=t^3y,\qquad q=t^3y^{-1} \, .
\ee
With these preparations, the building blocks~\eqref{consmet}  for the index can be written in the following compact form
\be\label{consmetell}
C_{\a_i\a_j\a_k}&=&\exp\left(\sum_{n=1}^\infty\frac{1}{n}\,g_n\;\chi_{3f}(n\a_i,n\a_j,n\a_k)\right)= \Gamma(\frac{t^2}{\sqrt{v}}a_i^{\pm1}a_j^{\pm1}a_k^{\pm1};p,q),\\
\eta^{\a_i}&=&\exp\left(\sum_{n=1}^\infty\frac{1}{n}\,f_n\; \chi_{adj}(n\a_i)\right)%\nonumber\\
= \frac{1}{\Delta(\a_i)}\,\frac{(p;p)(q;q)}{4\pi}\,\Gamma(t^2\, v;p,q)\,\frac{\Gamma(t^2\, v\,a_i^{\pm2};p,q)}{\Gamma(a_i^{\pm2};p,q)}\, .\nonumber
\ee 
Here we have defined $a_i=\exp(i\a_i)$ and used 
\be 
 \exp\left(\sum_{n=1}^\infty\frac{1}{n}\,f_n\right)&=&(p;p)(q;q)\,\Gamma(t^2\, v;p,q),\quad (a;b)\equiv\prod_{k=0}^\infty (1-a\,b^{k}) \,.
\ee  Again, the reader should keep in mind that  the {\it rhs} of the first line in~\eqref{consmetell} is a product of eight elliptic Gamma functions
according to the condensed notation~\eqref{condensed}.

\begin{table}
\begin{center}
\begin{tabular}{|m{0.5in}|m{1.3in}|c|} 
\hline 
\footnotesize{Symbol} & \footnotesize{Surface} & \footnotesize{Value} \\
\hline
\hline 
&&\\
$C_{\a\b\g}$ & \epsfig{file=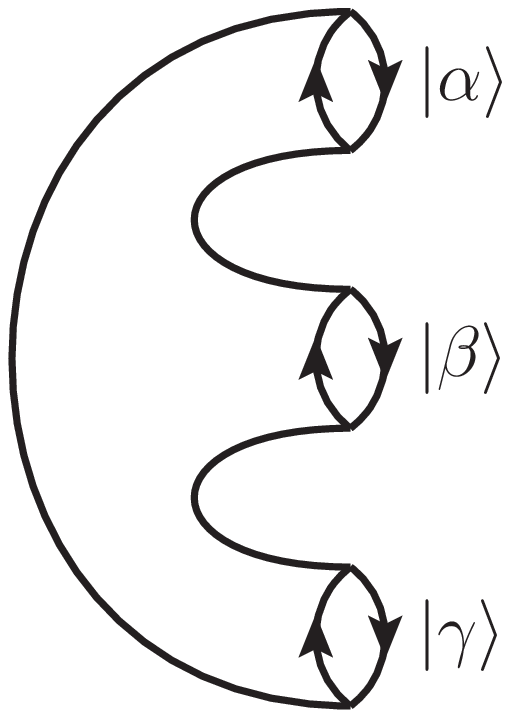,scale=0.4}  & $\Gamma(\frac{t^2}{\sqrt{v}}a^{\pm1}b^{\pm1}c^{\pm1})$ \\
%&&&&&\\
\hline 
&&\\
$C_{\a\b}^{\; \; \; \g}$ & \epsfig{file=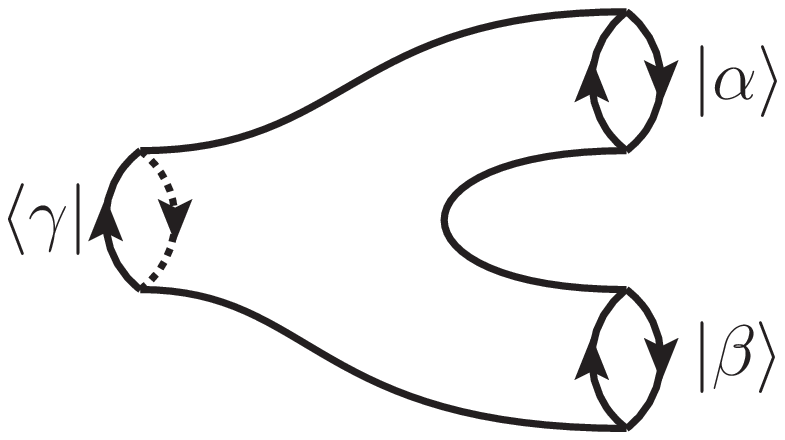,scale=0.4}  & $\frac{i\kappa}{\Delta(\g)}\,\Gamma(t^2\, v)\,\frac{\Gamma(t^2\, v\,c^{\pm2})}{\Gamma(c^{\pm2})}\Gamma(\frac{t^2}{\sqrt{v}}a^{\pm1}b^{\pm1}c^{\pm1})$ \\
%&&&&&\\
\hline
&&\\
$\eta^{\a\b}$ & \epsfig{file=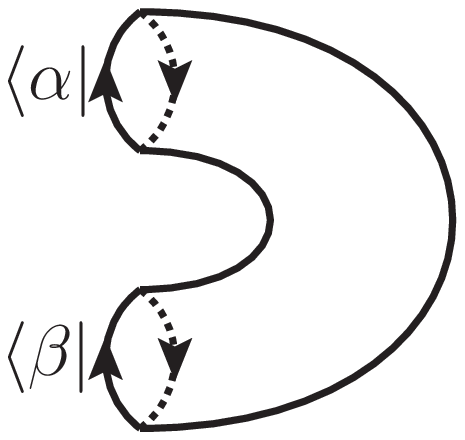,scale=0.4} & $\frac{i\,\kappa}{\Delta(\a)}\,\Gamma(t^2\, v)\,\frac{\Gamma(t^2\, v\,a^{\pm2})}{\Gamma(a^{\pm2})}\,\hat\delta(\a,\b)$\\
%&&&&&\\
\hline 
\end{tabular}
\end{center}
\caption{\label{strucs1}The structure constants and the metric in terms of elliptic Gamma functions.  For brevity 
we have left implicit the  parameters of the Gamma functions, $p= t^3 y$ and $q= t^3 y^{-1}$. We have
defined  $a\equiv \exp (i\a),\,b\equiv \exp (i\b)$, and $ c\equiv \exp (i \g)$. 
Recall also $\kappa \equiv (p;p)(q;q)/4\pi i$ and $\Delta(\alpha) \equiv (\sin^2 \alpha)/\pi$.  
 }
\end{table}

Collecting all these definitions  the fundamental integral  \eqref{indexS} becomes
\be\label{ellindex}
\kappa\,\Gamma\left(t^2v;p,q\right)\,\oint \frac{dz}{z}\;\frac{\Gamma(t^2\, v\,z^{\pm2};p,q)}{\Gamma(z^{\pm2};p,q)}\; \Gamma(\frac{t^2}{\sqrt{v}}a^{\pm1}b^{\pm1}z^{\pm1};p,q)
\;\Gamma(\frac{t^2}{\sqrt{v}}c^{\pm1}d^{\pm1}z^{\pm1};p,q),\quad pq=t^6\,,\nonumber\\
\ee with $\kappa \equiv (p;p)(q;q)/4\pi i$.
As it turns out, this integral fits into  a  class of integrals which are an active subject of mathematical research,  the elliptic Beta integrals
\be 
E^{(m)}(t_1,\dots,t_{2m+6})\sim\oint \frac{dz}{z}\frac{\Gamma(t_1 z,\dots t_{2m+6} z;p,q)}{\Gamma(z^{\pm 2};p,q)} \, ,
\qquad \prod_{k=1}^{2m+6}t_k=\left(pq\right)^{m+1} \, .
\ee  
Our integral is a special case of $E^{(5)}$. Elliptic Beta integrals  have very interesting
symmetry properties. For instance the symmetry of  $E^{(2)}$  is related to the Weyl group of $E_7$. Very recently van de Bult proved \cite{Focco}
that special cases of the $E^{(5)}$ integral, which are equivalent to \eqref{ellindex}, 
are invariant under the Weyl group of $F_4$. 
 In particular  \eqref{ellindex} is invariant under 
$b\leftrightarrow c$. This is theorem 3.2 in \cite{Focco},  with the parameters $\{t_{1,2,3,4},\, b\}$  of \cite{Focco} related to the parameters  $\{a,\,b,\,c,\,d,\, t^2v\}$ 
 in  our equation \eqref{ellindex} by the substitution
\be
&&t_1 \to \frac{t^2}{\sqrt{v}}\, a\, b ,\;\;
t_2 \to \frac{t^2}{\sqrt{v}}\, a / b,\;\;
t_3 \to \frac{t^2}{\sqrt{v}}\,c\, d ,\;
t_4 \to \frac{t^2}{\sqrt{v}}\, c/d,\;\;
b \to t^2\, v. 
\ee
This completes the proof of crossing symmetry  of the fundamental integral \eqref{indexS}.

\begin{figure}[h]
\begin{center}
\includegraphics[scale=0.45]{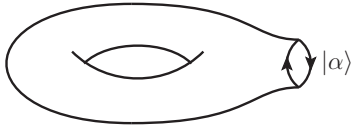}
\par\end{center}
\caption{\label{handle} Handle-creating operator ${\mathcal J}_\a$}
\end{figure}

The expressions for the structure constants and metric of the topological algebra in terms of the elliptic Gamma functions are summarized in Table~\ref{strucs1}. These expressions are analytic functions of  their arguments, except for for the metric $\eta^{\a\b}$ which
 contains a delta-function. One can try and use the results of the theory of elliptic Beta 
integrals to represent the delta-function in a more elegant way, indeed  
such a representation is sometimes available in terms of a contour integral~\cite{spirinv}. However, for generic choices
of the parameters, the definition of~\cite{spirinv} involves contour integrals not around the unit circle and thus using this representation
one presumably should also change the prescription~\eqref{contraction} for contracting indices.
In the limit $v\to t$ the relevant contours do approach the unit circle and the formalism of~\cite{spirinv} yields elegant expressions.
This limit is however slightly singular. We discuss it in Appendix \ref{tvapp}.

As a simple illustration of the use of the expressions in Table \ref{strucs1}
let us compute the superconformal index of the theory associated to  diagram {\bf (b)} in Figure \ref{degen}.
 This is essentially the ``handle-creating'' vertex ${\mathcal J}_\a$ of the TQFT, Figure~\ref{handle}.  We have
\be 
{\mathcal J}_\a=C_{\a\b\g}\,\eta^{\b\g}=\kappa\,\Gamma\left(t^2v\right)\,\Gamma\left(\frac{t^2}{\sqrt{v}} a^{\pm1}\right)^2\,\oint\frac{dz}{z}\frac{\Gamma(t^2v\, z^{\pm2})}{\Gamma(z^{\pm2})}
\Gamma\left(\frac{t^2}{\sqrt{v}} z^{\pm2}\,a^{\pm1}\right) \,.
\ee

Multivariate extensions  of elliptic Beta integrals 
have appeared in the calculation of the superconformal
index for pairs of  ${\cal N} = 1$ theories related by Seiberg duality~\cite{Dolan:2008qi}.
Unlike our ${\cal N} = 2$ superconformal cases, 
there is no continuous deformation relating two Seiberg-dual theories, and it is not a priori
obvious that their indices, evaluated at the free UV fixed points, should coincide --
but it turns out that they do,  thanks to identities satisfied by these multivariate integrals~\cite{rains}. 
See also~\cite{Spiridonov:2008zr}.  In Appendix \ref{n4app} we tackle the ${\cal N} = 4$ case,
evaluating the indices the S-dual pairs
with gauge  groups $Sp(n)$ and $SO(2n+1)$. Again S-duality  predicts 
some new identities of elliptic Beta integrals, which we  confirm  to the first few orders in the $t$ expansion.
It appears that there is a general connection between elliptic hypergeometric mathematics
and  electric-magnetic duality of the index of
$4d$ gauge theories.

\section{Discussion}

 A rich class of 4d superconformal field theories arise by
compactifying the 6d $(2,0)$ theory on a punctured Riemann surface $\Sigma$ \cite{Gaiotto:2009we}, and this has inspired
a precise dictionary between 4d  and 2d quantities  \cite{Alday:2009aq, Wyllard:2009hg, Drukker:2009tz, Drukker:2009id, Alday:2009fs}.
In this paper we have added a new entry to this dictionary. Previous work
has focussed on the relation between the 4d theory on $S^4$ (or more generally on the theory in the $\Omega$ background)
and Liouville theory on $\Sigma$.  Here we have considered instead the superconformal index \cite{Kinney:2005ej}, which can be viewed
as the partition function of the 4d theory on $S^3 \times S^1$, with  twisted boundary conditions 
labelled by three chemical potentials. We have argued that the superconformal index is evaluated
by a topological QFT on $\Sigma$. In the $A_1$ case we have computed explicitly the structure constants 
of the topological algebra and checked its associativity, using a rather non-trivial piece of contemporary mathematics \cite{Focco}.
Physically this result can be regarded as a precise check that the protected spectrum of operators is the same
for the $SU(2)^{N_G}$ theories related by the generalized S-dualities of \cite{Gaiotto:2009we}.

There are several interesting directions for future research. It would be illuminating to obtain a Lagrangian
description of the $2d$ TQFT from a twisted compactification of the $(2,0)$ theory on $S^3 \times S^1$,
and reproduce by that route the structure constants evaluated in this paper. 
The best known example of a topological field theory with observables labelled by the representations of $SU(2)$ is $2d$ Yang-Mills theory,
and it is likely that our theory will turn out to be related to it. There is then the related question
of finding  how this  structure can be embedded in string theory, perhaps along the lines of \cite{Dijkgraaf:2009pc}.
Finally  our  work should be extended to the $A_{N-1}$ theories with $N >2$. While for these theories a 4d Lagrangian
description is in general lacking, there are indirect ways to construct them  by taking  limits of known theories. The mathematical structure of the superconformal index is so rigid
that it may be possible to determine it by consistency, using purely 4d considerations.
 Alternatively, the ``top-down'' approach from compactification of the $(2,0)$ theory is expected to give a uniform answer for all  the $A_{N-1}$ theories.

We suspect that we are just scratching the surface of a general connection 
 between elliptic hypergeometric mathematics and S-duality. It is possible to generate new elliptic hypergeometric identities by calculating
 the superconformal index of S-dual theories.  Already the simplest S-dualities (from a physical perspective), such as the $SO(2n+1)/Sp(n)$
dualities in ${\cal N} = 4$ SYM, 
lead to  identities that to the best of our knowledge have not appeared in the mathematical literature. 
One may wonder whether the logic can be reversed, and new S-dualities discovered from known elliptic identities. 
Elliptic Beta integrals are the most general known extensions of the classic Euler Beta integral,
and as such they are the natural mathematical objects to appear in the calculation of ``crossing-symmetric''  physical quantities.
It is perhaps not coincidental that  the mathematics and  the physics  of the subject are being developed simultaneously,
and we can look forward to a fruitful interplay between the two viewpoints.

\section*{Acknowledgements}
We thank Fokko J. van de Bult and Eric Rains for very useful correspondence on elliptic Beta integrals and for 
comments on a draft of this paper, and Davide Gaiotto for useful discussions. 
This work was  supported in part by DOE grant DEFG-0292-ER40697 and by NSF grant PHY-0653351-001. Any
opinions, findings, and  conclusions or recommendations expressed in this 
material are those of the authors and do not necessarily reflect the views of the National 
Science Foundation.

\appendix

\section{S-duality for ${\mathcal N}=4$ $SO(2n+1)/Sp(n)$ SYM}\label{n4app}
In this Appendix we compute the superconformal indices  for  ${\mathcal N}=4$ SYM with gauge
groups $SO(2n+1)$ and $Sp(n)$. Since the $SO$ and $Sp$ theories are  related by S-duality, their  indices are  expected
to agree.   
These are in fact the only non-trivial ${\cal N} = 4$ cases from the viewpoint of index calculations.
 Indeed the index depends on the adjoint representation of the group:  the A, D, E, F and G cases are manifestly self-dual, and the only interesting duality is $B \leftrightarrow C$.

The characters of the adjoint representations of for $Sp(n)$ and $SO(2n+1)$ are 
\be
\chi_{Sp(n)}(\{z_i\})\quad:\quad &\sum_{1\leq i<j\leq n}\left(z_iz_j+z_iz_j^{-1}+z_jz_i^{-1}+z_i^{-1}z_j^{-1}\right)+
\sum_{i=1}^n(z_i^2+z_i^{-2})+n,\nonumber\\
\chi_{SO(2n+1)}(\{z_i\})\quad:\quad &  \sum_{1\leq i<j\leq n}\left(z_iz_j+z_iz_j^{-1}+z_jz_i^{-1}+z_i^{-1}z_j^{-1}\right)+
\sum_{i=1}^n(z_i+z_i^{-1})+n.\nonumber\\
\ee Their Haar measures   are 
\be 
Sp(n)\;&:&\; \int_{Sp(n)}d\mu(z)f(z)=\frac{(-)^n}{2^n\,n!}\oint_{{\mathbb T}_n}\prod_{j=1}^n\frac{dz_j}{2\pi i z_j}
\prod_{j=1}^n(z_j-z_j^{-1})^2\Delta({\bf z}+{\bf z}^{-1})^2\,f(z),\\
SO(2n+1)\;&:&\; \int_{So(2n+1)}d\mu(z)f(z)=\frac{(-)^n}{2^n\,n!}\oint_{{\mathbb T}_n}\prod_{j=1}^n\frac{dz_j}{2\pi i z_j}
\prod_{j=1}^n(z_j^{1/2}-z_j^{-1/2})^2\Delta({\bf z}+{\bf z}^{-1})^2\,f(z),\nonumber
\ee where ${\mathbb T}_n$ is an $n$-dimensional torus with unit radii and $\Delta({\bf x})$  the van der Monde determinant
\be 
\Delta({\bf x})=\prod_{i<j}(x_i-x_j) \,.
\ee The single letter partition function is  in both cases equal to~\cite{Kinney:2005ej}
\be
f(t,y)=\frac{3t^2-3t^4-t^3(y+y^{-1})+2t^6}{(1-t^3\,y)(1-t^3y^{-1})} \, ,
\ee where for simplicity we have  omitted the chemical potentials of the R-charges  -- we will restore them in the end.
 Using the identities \eqref{dolanid},
\be 
&&e^{\sum_k\frac{f_k}{k}\chi_{Sp(n)}(\{z_i^k\})}=\Gamma^{3n}(t^2;p,q)(p;p)^n(q;q)^n\prod_{i<j}\frac{z_j^2}{(1-z_iz_j)^2(1-z_i^{-1}z_j)^2}
\frac{1}{\Gamma(z_i^{\pm1}z_j^{\pm1};p,q)}\nonumber\\
&&\qquad\qquad\qquad\prod_{j}\frac{-z_j^2}{(1-z_j^2)^2}
\frac{1}{\Gamma(z_j^{\pm2};p,q)}\prod_{i<j}\Gamma(t^2z_i^{\pm1}z_j^{\pm1};p,q)^3\prod_{j}\Gamma(t^2z_i^{\pm2};p,q)^3.
\ee Recall the definition ~\eqref{consmetell} of the
 product $(x;y)$. Further, using
\be 
&&\prod_{i<j}(1-z_iz_j)(1-z_i/z_j)(1-z_j/z_i)(1-1/(z_iz_j))=\Delta({\bf z}+{\bf z}^{-1})^2,\\
&&\prod_j(1-z_j^2)(1-1/z_j^2)=(-1)^n\prod_j(z_j-1/z_j)^2\, ,\nonumber
\ee we obtain 
\be\label{sp}
 &&\int_{Sp(n)}d\mu(z)\,e^{\sum_k\frac{1}{k}f_k\chi_{Sp(n)}(\{z_i\})}=\\ &&\qquad\qquad
\frac{\Gamma^{3n}(t^2;p,q)}{2^n\,n!}(p;p)^n(q;q)^n
\oint\prod_j\frac{dz_j}{2\pi i z_j}\prod_{i<j}\frac{\Gamma(t^2z_i^{\pm1}z_j^{\pm1};p,q)^3}{\Gamma(z_i^{\pm1}z_j^{\pm1};p,q)}
\prod_j\frac{\Gamma(t^2z_j^{\pm2};p,q)^3}{\Gamma(z_j^{\pm2};p,q)}.\nonumber
\ee In complete analogy we obtain for the $SO(2n+1)$ gauge group
\be\label{so}
 &&\int_{SO(2n+1)}d\mu(z)\,e^{\sum_k\frac{1}{k}f_k\chi_{So(2n+1)}}=\\ &&\qquad\qquad
\frac{\Gamma^{3n}(t^2;p,q)}{2^n\,n!}(p;p)^n(q;q)^n
\oint\prod_j\frac{dz_j}{2\pi i z_j}\prod_{i<j}\frac{\Gamma(t^2z_i^{\pm1}z_j^{\pm1};p,q)^3}{\Gamma(z_i^{\pm1}z_j^{\pm1};p,q)}
\prod_j\frac{\Gamma(t^2z_j^{\pm1};p,q)^3}{\Gamma(z_j^{\pm1};p,q)}\,.\nonumber
\ee 
 S-duality predicts that the integrals (\ref{sp}) and (\ref{so}) must agree. For $Sp(1)\cong SO(3)$ this is trivially checked
by a change of variable:  in the $SO(3)$ integral make the substitution  $z\to y=\sqrt{z}$. The case of $Sp(2)\cong SO(5)$ is also
trivial (as it should be). Define $\hat z_1=\sqrt{z_1 z_2}$ and $\hat z_2=\sqrt{z_1 /z_2}$. Then in \eqref{so} the first product is 
exchanged with the second  with a doubled power of the $z$ argument and we obtain \eqref{sp}. We have checked for the first
few orders in  a series expansion  in $t$ that   \eqref{sp}  \eqref{so} also agree for higher rank groups.
 We do not have an analytic   proof of this statement.

Given an orthonormal  basis $e_i$ of ${\mathbb R}^n$ the root system of $C_n$ ($Sp(n)$)
consists of vectors of the form $X(C_n)=\{\pm2e_i,\,\pm e_i\pm e_j,i<j\}$. The root system of $B_n$ ($SO(2n+1)$) on the other hand
consists of vectors of the form  $X(B_n)=\{\pm e_i,\,\pm e_i\pm e_j,i<j\}$. These two systems are dual to one other.
The  integrands in \eqref{sp}
 and \eqref{so} are given by
\be
\prod_{\a\in X}\frac{\Gamma(t^2\,e^\a;p,q)^3}{\Gamma(e^\a;p,q)} \, ,
\ee where $X$ is the corresponding root system and we formally identify $z_i=e^{e_i}$. In this language 
it is easy to understand why the integrals \eqref{so} with $SO(3)/ SO(5)$, \eqref{sp} with $Sp(1)/Sp(2)$ are equal to one other.
In these cases the two root systems are linear transformations of one other, {\it i.e.} rescaling and in the case of $Sp(2)/SO(5)$ also rotation. 
For higher $n$ the relation is more complicated. For example for $n=3$ the $SO(7)$  lattice is a cube and the $Sp(3)$ lattice is
an octahedron.

Finally, let us indicate how the expressions for the indices are modified by adding the chemical 
potentials for the R-symmetry charges~\cite{Kinney:2005ej}. The only differences are in the numerators of 
(\ref{sp},\ref{so}), which become
\be
Sp(n):\quad &&\prod_{i<j}\Gamma(t^2 v\,z_i^{\pm1}z_j^{\pm1};p,q)\Gamma(\frac{t^2}{w} \,z_i^{\pm1}z_j^{\pm1};p,q)\Gamma(\frac{wt^2}{v} \,z_i^{\pm1}z_j^{\pm1};p,q)\\
&&\prod_j\Gamma(t^2v\,z_j^{\pm2};p,q)\Gamma(\frac{t^2}{w}\,z_j^{\pm2};p,q)\Gamma(\frac{wt^2}{v}\,z_j^{\pm2};p,q),\nonumber\\
SO(2n+1):\quad &&\prod_{i<j}\Gamma(t^2 v\,z_i^{\pm1}z_j^{\pm1};p,q)\Gamma(\frac{t^2}{w} \,z_i^{\pm1}z_j^{\pm1};p,q)\Gamma(\frac{wt^2}{v} \,z_i^{\pm1}z_j^{\pm1};p,q)\nonumber\\
&&\prod_j\Gamma(t^2v\,z_j^{\pm1};p,q)\Gamma(\frac{t^2}{w}\,z_j^{\pm1};p,q)\Gamma(\frac{wt^2}{v}\,z_j^{\pm1};p,q)\, ,\nonumber
\ee and in the prefactor of the integrals,
\be
\Gamma^{3n}(t^2;p,q)\quad\to\quad \Gamma^{n}(t^2\,v;p,q)\,\Gamma^{n}(\frac{t^2}{w};p,q)\,\Gamma^{n}(\frac{w\,t^2}{v};p,q)\,.
\ee

\section{The Representation Basis}\label{repbasis}

The labels of the topological algebra as we have defined in \eqref{consmet} are (compact) continuous parameters $\alpha_i \in [0, 2 \pi)$.
 We can ``Fourier'' transform to the discrete basis  of irreducible $SU(2)$ representations.
We denote  by $R_K$ the  irreducible representation of $SU(2)$ of dimension $K+1$. 
The integrals over characters translate into  sums over representations.  The structure constants in the discrete
basis are given by
\be\label{vertfour}
C_{\a\b\g}&=&\sum_{K,L,M=0}^\infty \frac{\sin (K+1)\a}{\sin\a}\,\frac{\sin (L+1)\b}{\sin\b}\, \frac{\sin (M+1)\g}{\sin\g}\;\hat C_{KLM}\\
&=&\sum_{K,L,M=0}^\infty \chi_K(\a)\chi_L(\b)\chi_M(\g)\;\hat C_{KLM},\nonumber
\ee where $\chi_K(\a)$ is the character of $R_K$,
\be
\chi_K(\a)=\frac{\sin (K+1)\,\a}{\sin\a}\,. 
\ee 
Similarly the metric in the discrete basis is given by
\be
\eta^{\a\b}&=&\sum_{K,\,L\,=\,0}^\infty \chi_K(\a)\chi_L(\b)\;\hat \eta^{KL} \,.
\ee 
Further, we define the scalar product of characters\footnote{We have a slightly different
convention for the characters and thus the expression of the scalar product differs from the one  in~\cite{Dolan:2007rq}.}
\be\label{scalarprod}
&&\langle\chi_K\,\chi_M\rangle=\frac{1}{2\pi i}\oint \frac{dz}{z}\,(1-z^2)\,\chi_K(z)\,\chi_M(z)\\
&&\quad\qquad=-\frac{1}{4\pi i}\oint \frac{dz}{z}\,(z-\frac{1}{z})^2\,\chi_K(z)\,\chi_M(z)=
\int_0^{2\pi}d\theta\Delta(\theta)\chi_K(\theta)\chi_M(\theta)=\delta_{K,M}\, .\nonumber
\ee In the second equality we have introduced the  measure \eqref{measure} and used the fact that $\chi(z)=\chi(z^{-1})$. 
Thus we have 
\be
\sum_{K=0}^\infty \chi_K(\a)\,\chi_K(\b)=\hat\delta(\a,\b),\qquad \int_0^{2\pi}d\theta\Delta(\theta)\, \hat\delta(\theta,\a)\, f(\theta)=f(\a),  
\ee for any $f$ obeying $f(\theta)=f(-\theta)$. Using \eqref{consmet} we can write
\be
\hat\eta^{KL} = \eta^I\langle\chi^I\chi^K\chi^L\rangle,\qquad \eta^I=\int d\a\,\Delta(\a)\,\eta^\a\,\chi_I(\a).
\ee 

 \noindent Finally with the help of  these definitions, we can rewrite \eqref{topindex} as
\be\label{topindexdisc}
{\mathcal I}=\prod_{\{i,j,k\}\in {\mathcal V}} \hat C_{L_i L_j L_k}\,\prod_{\{m,n\}\in {\mathcal G}}\hat \eta^{L_m L_n}\,,
\ee where  index contractions  now indicate  sums over the non-negative integers.

%%%%%%%%%%%%%%

\section{TQFT Algebra for $v = t$}\label{tvapp}

For $v = t$ we can rewrite the algebra of the topological quantum field theory~\eqref{consmet} in a more elegant way,
removing the delta-functions by making use of  identities obeyed by elliptic Beta integrals. 
This does  not appear to be a  preferred limit  physically, except for the fact that the contribution to the index of the chiral superfield
in the ${\mathcal N}=2$ vector multiplet vanishes, see \eqref{letterpart}.
Our manipulations
will be slightly formal since the limit $v=t$ of the formulae we will use is somewhat singular.
We start by quoting  the  important identity
\be\label{BetaEll}
E^{(m=0)}(t_1,\dots,t_6)&=&\kappa \oint \frac{dz}{z}\frac{\prod_{k=1}^6\Gamma\left(t_k\,z^{\pm1};p,q\right)}{\Gamma\left(z^{\pm2};p,q\right)}=
\prod_{1\leq j<k\leq 6}\Gamma\left(t_j\,t_k;p,q\right),\qquad \prod_{k=1}^6t_k=pq \, .\nonumber\\
\ee This is a vast generalization to elliptic Gamma functions of that seminal object in string theory, the classic Beta integral of Euler,
\be\label{Beta}
B(\a,\b)=\int_0^1dt\,t^{\a-1}(1-t)^{\b-1}=\frac{\Gamma(\a)\Gamma(\b)}{\Gamma(\a+\b)} \, ,
\ee 
which is recovered as a special limit, see {\it e.g}.~\cite{Spiridonov}.
Applying \eqref{BetaEll} we have
\be\label{BetaEll1}
&&\kappa\oint \frac{dz}{z}\frac{\Gamma\left(\tau \sqrt{\nu}\,a^{\pm1}b^{\pm1}z^{\pm1}\right)\,\Gamma\left(\frac{\tau}{\nu}\,z^{\pm1}y^{\pm1}\right)}
{\Gamma\left(z^{\pm2}\right)}= \\
&&\qquad \qquad \Gamma\left(\frac{\tau^2}{\sqrt{\nu}}\,a^{\pm1}b^{\pm1}y^{\pm1}\right)\;
\Gamma\left(\tau^2 \nu\,a^{\pm2}\right)\Gamma\left(\tau^2 \nu \,b^{\pm2}\right)\,\Gamma\left(\frac{\tau^2}{\nu^2}\right)\,\Gamma\left(\tau^2\, \nu \right)^2\,.\nonumber
\ee 
For  brevity we have omitted the $p$ and $q$ parameters in the Gamma functions. We assume $pq = \tau^6$. For these values of $p$ and $q$, 
$\Gamma(\tau^3z^{\pm1})=1$. Now if we take $\nu = \tau$, 
\be \label{formaldelta}
\kappa\oint \frac{dz}{z}\frac{\Gamma\left(\tau^{3/2}\,a^{\pm1}b^{\pm1}z^{\pm1}\right)\,\Gamma\left(z^{\pm1}y^{\pm1}\right)}
{\Gamma\left(z^{\pm2}\right)}= 
 \Gamma\left(\tau^{3/2} \,a^{\pm1}b^{\pm1}y^{\pm1}\right)\;
\Gamma\left(1 \right) \,.
\ee 
Strictly speaking the elliptic Beta integral formula~\eqref{BetaEll} holds when  $|t_k|<1$ for all $k=1\dots6$. For
$\nu = \tau$ some of the $t_k$s in~\eqref{BetaEll1} saturate this bound. The elliptic Beta integral~\eqref{BetaEll1} is proportional to $\Gamma(\frac{\tau^2}{\nu^2};p,q)\to\Gamma(1;p,q)$. Since
the elliptic Gamma function has a simple pole when its argument approaches $z=1$ (see~\eqref{gamma1}), \eqref{BetaEll1} diverges in the limit.
We will proceed by keeping formal factors of $\Gamma(1)$ in all the expressions. Thanks to (\ref{formaldelta}), the expression
\be
\frac{\Gamma(z^{\pm 1} y^{ \pm 1} )}{\Gamma(z^{\pm 2}) \Gamma(1)} \equiv \delta^z_y
\ee
acts as a formal identity operator.
 All factors of $\Gamma(1)$ will cancel in the final expression for the index.

\begin{table}[th]
\begin{center}
\begin{tabular}{|m{0.5in}|m{0.9in}|l||m{0.5in}|m{1.1in}|l|} 
\hline 
\footnotesize{Symbol} & \footnotesize{Surface} & \footnotesize{Value} & \footnotesize{Symbol} & \footnotesize{Surface} & \footnotesize{Value}\\
\hline
\hline 
&&&&&\\
$C_{abc}$ & \epsfig{file=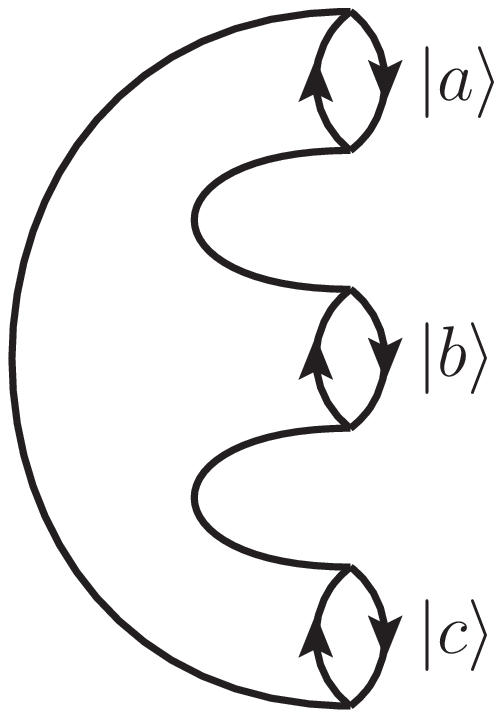,scale=0.4}  & $\Gamma(t^{\frac{3}{2}}a^{\pm1}b^{\pm1}c^{\pm1})$ & $V^{a}$ & \epsfig{file=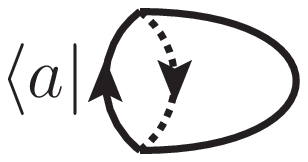,scale=0.4}  & $\frac{1}{\Gamma(1)^2}\,\frac{\Gamma(t^{\pm\frac{3}{2}}a^{\pm1})}{\Gamma(a^{\pm2})}$\\
%&&&&&\\
\hline
&&&&&\\
$\eta^{ab}$ & \epsfig{file=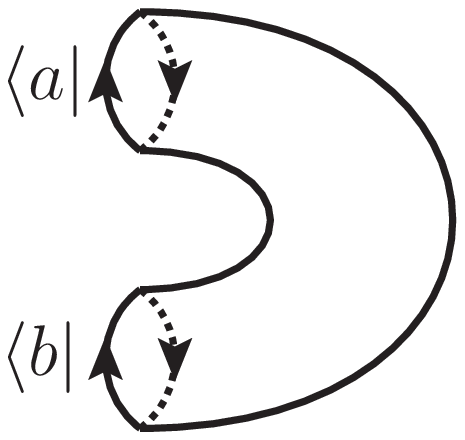,scale=0.4} & $\frac{1}{\Gamma(1)}\,\frac{\Gamma(a^{\pm1}b^{\pm1})}{\Gamma(a^{\pm2},b^{\pm2})}$& $\eta_{ab}$ & \epsfig{file=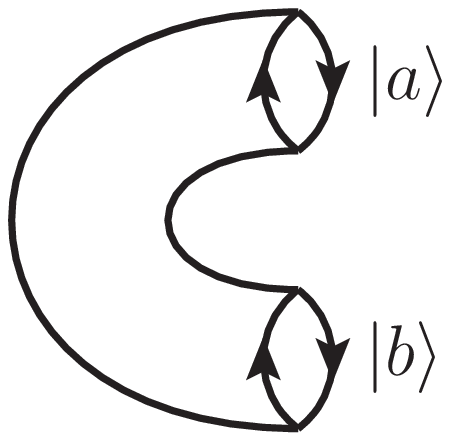,scale=0.4} & $\frac{1}{\Gamma(1)}\Gamma\left(a^{\pm1}b^{\pm1}\right)$\\
%&&&&&\\
\hline 
\end{tabular}
\end{center}
\caption{\label{tab:buildingblocks}The basic building blocks of the topological algebra in the $v = t$ case.}
\end{table}

\begin{figure}[htbp]
\begin{center}
\includegraphics[scale=0.4]{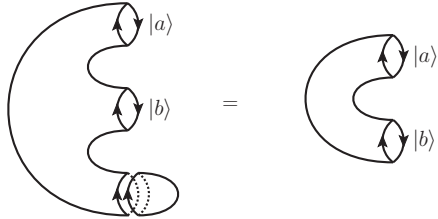}
\par\end{center}
\caption{\label{fig:cappingoff}Constructing the metric by capping
off the trivalent vertex.}
\end{figure}

%\noindent 
For $t=v$ we can write the building blocks of the topological algebra in the form summarized in Table \ref{tab:buildingblocks}. 
Contraction of indices is defined as
\be\label{integind}
A^{..a..}\,B_{..a..}\to \kappa\,\oint\frac{d\,a}{a}\,A^{..a..}\,B_{..a..}. 
\ee 
We now proceed to perform a few sample calculations and consistency checks.
We can raise an index of the structure constants to obtain
\be
C_{abe}\,\eta^{ec}=\frac{\kappa}{\Gamma(1)}\,
\oint\frac{d\,e}{e}\Gamma(t^{\frac{3}{2}}a^{\pm1}b^{\pm1}e^{\pm1})
\frac{\Gamma(e^{\pm1}c^{\pm1})}{\Gamma(e^{\pm2},c^{\pm2})}=\frac{\Gamma(t^{\frac{3}{2}}a^{\pm1}b^{\pm1}c^{\pm1})}{\Gamma(c^{\pm2})}
={C_{ab}}^c \, .\nonumber\\ 
\ee In particular we see that the index \eqref{ellindex} is finite and is simply given by ${C_{ab}}^c\,C_{cde}$.
The ``vacuum state'' $| V  \rangle \equiv V^a |a \rangle$ satisfies  
by definition (see {\it e.g.} \cite{Dijkgraaf:1997ip}) $C_{abc}\,V^c = \eta_{ab}$, as illustrated in Figure \ref{fig:cappingoff}.
This determines $V^a$ to be the expression in Table \ref{tab:buildingblocks}, 
\be 
C_{abc}\,V^c=\frac{\kappa}{\Gamma(1)^2}\oint\frac{dz}{z}\Gamma(t^{\frac{3}{2}}a^{\pm1}b^{\pm1}z^{\pm1})\,\frac{\Gamma(t^{\pm\frac{3}{2}}z^{\pm1})}{\Gamma(z^{\pm2})}=
\frac{1}{\Gamma(1)}\Gamma(a^{\pm1}b^{\pm1})=\eta_{ab}\,.
\ee 

\noindent Further, we can check that $\eta_{ab}$ and $\eta^{ab}$ in Table \ref{tab:buildingblocks} are one the inverse of the other,
\be
\eta^{ae}\,\eta_{ec}=\frac{\kappa}{\Gamma(1)^2}\oint\frac{d\,e}{e}\,
\frac{\Gamma(a^{\pm1}e^{\pm1})}{\Gamma(a^{\pm2},e^{\pm2})}\,\Gamma\left(e^{\pm1}c^{\pm1}\right)=
\frac{1}{\Gamma(1)}\,\frac{\Gamma(a^{\pm1}c^{\pm1})}{\Gamma(a^{\pm2})}=\delta^a_c \,.
\ee
\begin{figure}[htbp]
\begin{center}
\includegraphics[scale=0.4]{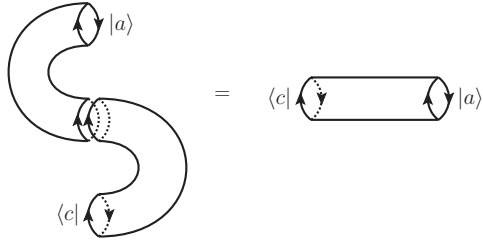}
\par\end{center}
\caption{\label{fig:inversemetric}Topological interpretation of the property $\eta^{ce}\,\eta_{ea}=\delta^{c}_{a}$.}
\end{figure}

\begin{figure}[htbp]
\begin{center}
\includegraphics[scale=0.4]{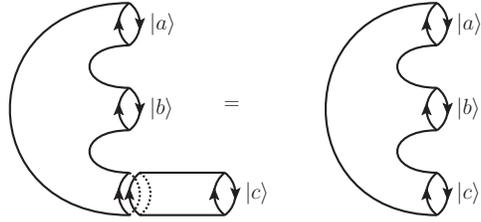}
\par\end{center}
\caption{\label{fig:deltacheck-C}The consistency requirement
$\delta_{c}^{z}\,C_{abz}=C_{abc}$.}
\end{figure}

\noindent As a consistency check one can verify in examples that $\delta^a_b$ is indeed an identity.
For instance 
\be
\delta^{z}_{a}\,C_{zbc}=\frac{\kappa}{\Gamma(1)}\oint\frac{dz}{z}\,\frac{\Gamma(a^{\pm1}z^{\pm1})}{\Gamma(z^{\pm2})}\,\Gamma(t^{\frac{3}{2}}z^{\pm1}b^{\pm1}c^{\pm1})=\Gamma(t^{\frac{3}{2}}a^{\pm1}b^{\pm1}c^{\pm1})=C_{abc} \, ,
\ee as illustrated in Figure \ref{fig:deltacheck-C}.
For completeness we can also compute the sphere and the torus partition functions.
(These partition functions  do not appear  in any index computation of  a 4d superconformal theory so their physical interpretation is unclear.)

\begin{figure}[htbp]
\begin{center}
$\begin{array}{c@{\hspace{0.65in}}c}
\epsfig{file=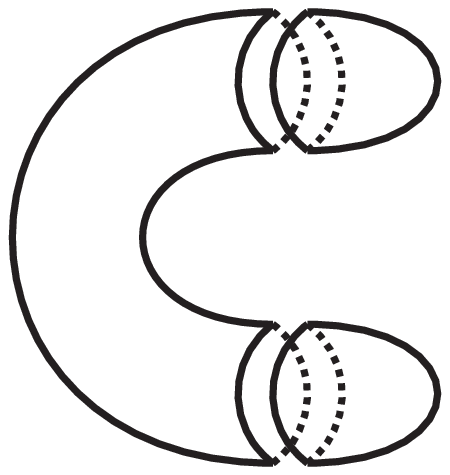,scale=0.40} & \epsfig{file=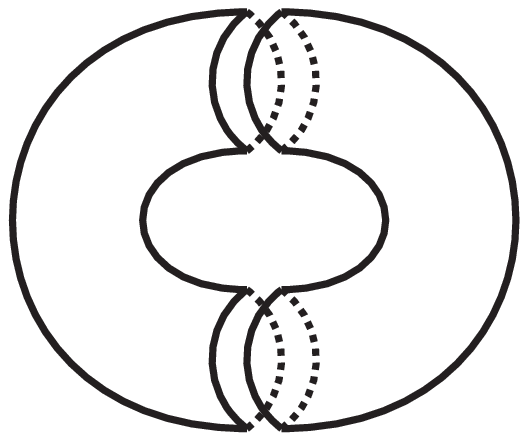,scale=0.4}\\\
(a) & (b) \\ [0.2cm]
\end{array}$
\end{center}
 \begin{center}
\caption{The sphere {\bf{(a)}} and the torus {\bf{(b)}} partition functions.
} \label{partitions}
\end{center}
\end{figure}

The sphere partition function is  given by
\be
V^c\,V^e\,\eta_{ce}&=&\frac{\kappa^2}{\Gamma(1)^5}\oint\frac{de}{e} \oint\frac{dc}{c}
\frac{\Gamma\left(c^{\pm1}e^{\pm1}\right)
\Gamma\left(t^{\pm3/2}\,c^{\pm1}\right)\,\Gamma\left(t^{\pm3/2}\,e^{\pm1}\right)}{\Gamma\left(c^{\pm2}\right)\Gamma\left(e^{\pm2}\right)}=\\
&=&\frac{\kappa}{\Gamma(1)^4}\oint\frac{de}{e} 
\frac{
\Gamma\left(t^{\pm3/2}\,e^{\pm1}\right)^2}{\Gamma\left(e^{\pm2}\right)}=\Gamma(t^{-3})\,\frac{1}{\Gamma(1)}.\nonumber
\ee The torus partition function is given by
\be
\eta_{ab}\eta^{ab}=\frac{\kappa}{\Gamma(1)}\,\oint\frac{d\,a}{a}\frac{\Gamma(a^{\pm1}a^{\pm1})}{\Gamma(a^{\pm2})}=
\kappa\,\Gamma(1)\,\oint\frac{d\,a}{a}=2\,\pi\, i\,\kappa\,\Gamma(1).
\ee 
Since $\Gamma(1) = \infty$  the sphere partition function vanishes and the torus partition function  
diverges.

%\newpage

\bibliography{sduality}

\providecommand{\href}[2]{#2}\begingroup\raggedright\begin{thebibliography}{10}

\bibitem{Gaiotto:2009we}
D.~Gaiotto, {\it {N=2 dualities}},
  \href{http://xxx.lanl.gov/abs/arXiv:0904.2715}{{\tt arXiv:0904.2715}}.

\bibitem{Focco}
F.~J. van~de Bult, {\it An elliptic hypergeometric integral with $w(f_4)$
  symmetry},  \href{http://xxx.lanl.gov/abs/arXiv:0909.4793}{{\tt
  arXiv:0909.4793}}.

\bibitem{Witten:2009at}
E.~Witten, {\it {Geometric Langlands From Six Dimensions}},
  \href{http://xxx.lanl.gov/abs/arXiv:0905.2720}{{\tt arXiv:0905.2720}}.

\bibitem{Nekrasov:2002qd}
N.~A. Nekrasov, {\it {Seiberg-Witten Prepotential From Instanton Counting}},
  {\em Adv. Theor. Math. Phys.} {\bf 7} (2004) 831--864,
  [\href{http://xxx.lanl.gov/abs/hep-th/0206161}{{\tt hep-th/0206161}}].

\bibitem{Alday:2009aq}
L.~F. Alday, D.~Gaiotto, and Y.~Tachikawa, {\it {Liouville Correlation
  Functions from Four-dimensional Gauge Theories}},
  \href{http://xxx.lanl.gov/abs/arXiv:0906.3219}{{\tt arXiv:0906.3219}}.

\bibitem{Wyllard:2009hg}
N.~Wyllard, {\it {$A_{N-1}$ conformal Toda field theory correlation functions
  from conformal N=2 SU(N) quiver gauge theories}},
  \href{http://xxx.lanl.gov/abs/arXiv:0907.2189}{{\tt arXiv:0907.2189}}.

\bibitem{Drukker:2009tz}
N.~Drukker, D.~R. Morrison, and T.~Okuda, {\it {Loop operators and S-duality
  from curves on Riemann surfaces}},  {\em JHEP} {\bf 09} (2009) 031,
  [\href{http://xxx.lanl.gov/abs/arXiv:0907.2593}{{\tt arXiv:0907.2593}}].

\bibitem{Drukker:2009id}
N.~Drukker, J.~Gomis, T.~Okuda, and J.~Teschner, {\it {Gauge Theory Loop
  Operators and Liouville Theory}},
  \href{http://xxx.lanl.gov/abs/arXiv:0909.1105}{{\tt arXiv:0909.1105}}.

\bibitem{Alday:2009fs}
L.~F. Alday, D.~Gaiotto, S.~Gukov, Y.~Tachikawa, and H.~Verlinde, {\it {Loop
  and surface operators in N=2 gauge theory and Liouville modular geometry}},
  \href{http://xxx.lanl.gov/abs/arXiv:0909.0945}{{\tt arXiv:0909.0945}}.

\bibitem{Mironov:2009qn}
A.~Mironov and A.~Morozov, {\it {Proving AGT relations in the large-c limit}},
  \href{http://xxx.lanl.gov/abs/arXiv:0909.3531}{{\tt arXiv:0909.3531}}.

\bibitem{Alday:2009qq}
L.~F. Alday, F.~Benini, and Y.~Tachikawa, {\it {Liouville/Toda central charges
  from M5-branes}},  \href{http://xxx.lanl.gov/abs/arXiv:0909.4776}{{\tt
  arXiv:0909.4776}}.

\bibitem{Poghossian:2009mk}
R.~Poghossian, {\it {Recursion relations in CFT and N=2 SYM theory}},
  \href{http://xxx.lanl.gov/abs/arXiv:0909.3412}{{\tt arXiv:0909.3412}}.

\bibitem{Marshakov:2009gn}
A.~Marshakov, A.~Mironov, and A.~Morozov, {\it {On non-conformal limit of the
  AGT relations}},  \href{http://xxx.lanl.gov/abs/arXiv:0909.2052}{{\tt
  arXiv:0909.2052}}.

\bibitem{Bonelli:2009zp}
G.~Bonelli and A.~Tanzini, {\it {Hitchin systems, N=2 gauge theories and
  W-gravity}},  \href{http://xxx.lanl.gov/abs/arXiv:0909.4031}{{\tt
  arXiv:0909.4031}}.

\bibitem{Marshakov:2009kj}
A.~Marshakov, A.~Mironov, and A.~Morozov, {\it {Zamolodchikov asymptotic
  formula and instanton expansion in N=2 SUSY $N_f=2N_c$ QCD}},
  \href{http://xxx.lanl.gov/abs/arXiv:0909.3338}{{\tt arXiv:0909.3338}}.

\bibitem{Nanopoulos:2009au}
D.~Nanopoulos and D.~Xie, {\it {On Crossing Symmmetry and Modular Invariance in
  Conformal Field Theory and S Duality in Gauge Theory}},
  \href{http://xxx.lanl.gov/abs/arXiv:0908.4409}{{\tt arXiv:0908.4409}}.

\bibitem{Gaiotto:2009ma}
D.~Gaiotto, {\it {Asymptotically free N=2 theories and irregular conformal
  blocks}},  \href{http://xxx.lanl.gov/abs/arXiv:0908.0307}{{\tt
  arXiv:0908.0307}}.

\bibitem{Gaiotto:2009gz}
D.~Gaiotto and J.~Maldacena, {\it {The gravity duals of N=2 superconformal
  field theories}},  \href{http://xxx.lanl.gov/abs/arXiv:0904.4466}{{\tt
  arXiv:0904.4466}}.

\bibitem{Tachikawa:2009rb}
Y.~Tachikawa, {\it {Six-dimensional $D_N$ theory and four-dimensional SO-USp
  quivers}},  {\em JHEP} {\bf 07} (2009) 067,
  [\href{http://xxx.lanl.gov/abs/arXiv:0905.4074}{{\tt arXiv:0905.4074}}].

\bibitem{Dijkgraaf:2009pc}
R.~Dijkgraaf and C.~Vafa, {\it {Toda Theories, Matrix Models, Topological
  Strings, and N=2 Gauge Systems}},
  \href{http://xxx.lanl.gov/abs/arXiv:0909.2453}{{\tt arXiv:0909.2453}}.

\bibitem{Benini:2009gi}
F.~Benini, S.~Benvenuti, and Y.~Tachikawa, {\it {Webs of five-branes and N=2
  superconformal field theories}},  {\em JHEP} {\bf 09} (2009) 052,
  [\href{http://xxx.lanl.gov/abs/arXiv:0906.0359}{{\tt arXiv:0906.0359}}].

\bibitem{Maruyoshi:2009uk}
K.~Maruyoshi, M.~Taki, S.~Terashima, and F.~Yagi, {\it {New Seiberg Dualities
  from N=2 Dualities}},  {\em JHEP} {\bf 09} (2009) 086,
  [\href{http://xxx.lanl.gov/abs/arXiv:0907.2625}{{\tt arXiv:0907.2625}}].

\bibitem{Nanopoulos:2009xe}
D.~Nanopoulos and D.~Xie, {\it {N=2 SU Quiver with USP Ends or SU Ends with
  Antisymmetric Matter}},  {\em JHEP} {\bf 08} (2009) 108,
  [\href{http://xxx.lanl.gov/abs/arXiv:0907.1651}{{\tt arXiv:0907.1651}}].

\bibitem{Benini:2009mz}
F.~Benini, Y.~Tachikawa, and B.~Wecht, {\it {Sicilian gauge theories and N=1
  dualities}},  \href{http://xxx.lanl.gov/abs/arXiv:0909.1327}{{\tt
  arXiv:0909.1327}}.

\bibitem{Kinney:2005ej}
J.~Kinney, J.~M. Maldacena, S.~Minwalla, and S.~Raju, {\it {An index for 4
  dimensional super conformal theories}},  {\em Commun. Math. Phys.} {\bf 275}
  (2007) 209--254, [\href{http://xxx.lanl.gov/abs/hep-th/0510251}{{\tt
  hep-th/0510251}}].

\bibitem{Spiridonov}
J.~van Diejen and V.~Spiridonov, {\it {Elliptic Beta Integrals and Mudular
  Hypergeometric Sums: An Overview}},  {\em Rocky Mountain J. Math.} {\bf 32
  (2)} (2002) 639--656.

\bibitem{Spiridonov2}
V.~Spiridonov, {\it {Essays on the theory of elliptic hypergeometric
  functions}},  {\em Uspekhi Mat. Nauk} {\bf 63 no 3} (2008) 3--72,
  [\href{http://xxx.lanl.gov/abs/arXiv:0805.3135}{{\tt arXiv:0805.3135}}].

\bibitem{Spiridonov3}
V.~Spiridonov, {\it {Classical elliptic hypergeometric functions and their
  applications }},  {\em Rokko Lect. in Math. Vol. 18, Dept. of Math, Kobe
  Univ.} (2005) 253--287,
  [\href{http://xxx.lanl.gov/abs/arXiv:math/0511579}{{\tt
  arXiv:math/0511579}}].

\bibitem{Dolan:2008qi}
F.~A. Dolan and H.~Osborn, {\it {Applications of the Superconformal Index for
  Protected Operators and q-Hypergeometric Identities to N=1 Dual Theories}},
  {\em Nucl. Phys.} {\bf B818} (2009) 137--178,
  [\href{http://xxx.lanl.gov/abs/arXiv:0801.4947}{{\tt arXiv:0801.4947}}].

\bibitem{Romelsberger:2007ec}
C.~Romelsberger, {\it {Calculating the Superconformal Index and Seiberg
  Duality}},  \href{http://xxx.lanl.gov/abs/0707.3702}{{\tt 0707.3702}}.

\bibitem{Dolan:2002zh}
F.~A. Dolan and H.~Osborn, {\it {On short and semi-short representations for
  four dimensional superconformal symmetry}},  {\em Ann. Phys.} {\bf 307}
  (2003) 41--89, [\href{http://xxx.lanl.gov/abs/hep-th/0209056}{{\tt
  hep-th/0209056}}].

\bibitem{atiyah}
M.~Atiyah, {\it {Topological quantum field theories}},  {\em Inst. Hautes
  Etudes Sci. Publ. Math.} {\bf 68} (1989) 175--186.

\bibitem{Dijkgraaf:1997ip}
R.~Dijkgraaf, {\it {Fields, strings and duality}},
  \href{http://xxx.lanl.gov/abs/hep-th/9703136}{{\tt hep-th/9703136}}.

\bibitem{spirinv}
S.~O. Warnaar and V.~Spiridonov, {\it {Inversions of integral operators and
  elliptic beta integrals on root systems }},  {\em Adv. Math.} {\bf 207}
  (2006) 91--132, [\href{http://xxx.lanl.gov/abs/math/0411044}{{\tt
  math/0411044}}].

\bibitem{rains}
E.~M. Rains, {\it {Transformations of Elliptic Hypergeometric Integrals }},
  \href{http://xxx.lanl.gov/abs/math/0309252}{{\tt math/0309252}}.

\bibitem{Spiridonov:2008zr}
V.~P. Spiridonov and G.~S. Vartanov, {\it {Superconformal indices for
  ${\mathcal N}=1$ theories with multiple duals}},
  \href{http://xxx.lanl.gov/abs/arXiv:0811.1909}{{\tt arXiv:0811.1909}}.

\bibitem{Dolan:2007rq}
F.~A. Dolan, {\it {Counting BPS operators in N=4 SYM}},  {\em Nucl. Phys.} {\bf
  B790} (2008) 432--464, [\href{http://xxx.lanl.gov/abs/arXiv:0704.1038}{{\tt
  arXiv:0704.1038}}].

\end{thebibliography}\endgroup
\bibliographystyle{JHEP}
%\bibliographystyle{unsrt}
%\bibliography{ref}
%\bibliographystyle{apsrev}
%\bibliographystyle{plain}
%\bibliographystyle{utphys}

\end{document}